\documentclass[twocolumn]{autart}    

\usepackage{graphicx}                           
\usepackage{mathrsfs}
\usepackage{graphicx}
\usepackage{subfigure}
\usepackage{enumerate}
\usepackage{url}
\usepackage{appendix}
\usepackage{amsmath}
\usepackage{comment}
\usepackage{color}
\usepackage{array}
\usepackage{dsfont}
\usepackage{amsfonts}
\usepackage{color}
\usepackage{flushend}
\usepackage{amssymb}
\usepackage{multicol} 
\usepackage{cite} 
\usepackage{subfigure}

\usepackage[noend]{algpseudocode}

\usepackage{algorithmicx,algorithm}

\newtheorem{theorem}{Theorem}
\newtheorem{proposition}{Proposition}

\newtheorem{definition}{Definition}
\newtheorem{lemma}{Lemma}

\newtheorem{remark}{Remark}
\newtheorem{example}{Example}

\begin{document}

\begin{frontmatter} 

\title{A  Framework for Current-State Opacity\\ under Dynamic Information Release Mechanism} %

\thanks[footnoteinfo]{This work was supported  by the National Key R\&D Program (2018AAA0101700),  the National Natural Science Foundation of China (61803259, 61833012), and the Shanghai Jiao Tong University Scientific and Technological Innovation Funds.  Corresponding author X.~Yin.}

\author[SJTU1]{Junyao Hou}\ead{houjunyao@sjtu.edu.cn},      
\author[SJTU1]{Xiang Yin}\ead{yinxiang@sjtu.edu.cn},               
\author[SJTU1]{Shaoyuan Li}\ead{syli@sjtu.edu.cn}

\address[SJTU1]{Department of Automation and Key Laboratory of System Control and Information Processing, Shanghai Jiao Tong University, Shanghai 200240, China.}

\begin{keyword}                           
Opacity, Discrete-Event Systems, Security, Information Release         
\end{keyword}                             

\begin{abstract}                           
Opacity is an important information-flow security property that characterizes the plausible deniability of a dynamic system for its ``secret" against eavesdropping attacks. As an information-flow property, the underlying observation model is the key in the modeling and analysis of opacity.  In this paper, we investigate the verification of current-state opacity for discrete-event systems under Orwellian-type observations, i.e., the system is allowed to re-interpret the observation of an event based on its future suffix.  First, we propose a new  Orwellian-type observation model  called the dynamic information release mechanism (DIRM). In the DIRM,  when to release previous ``hold on" events is state-dependent. 
Then we propose a new definition of opacity based on the notion of history-equivalence rather than the standard projection-equivalence. This definition is more suitable for observations that are not prefix-closed. 
Finally, we show that by constructing a new structure called the  DIRM-observer, current-state opacity can be effectively verified under the DIRM.  
Computational complexity analysis as well as illustrative examples for the proposed approach are also provided. 
Compared with the existing Orwellian-type observation model, the proposed framework is more general in the sense that the information-release-mechanism is state-dependent, information is partially released and the corresponding definition  of opacity is more suitable for non-prefix-closed observations. 
\end{abstract}

\end{frontmatter}

\section{Introduction} 
Security and privacy issues  are becoming pervasive in safety-critical cyber-physical systems as  computational devices nowadays are connected by networks which may lead to information leakage.  
For dynamic systems, an important angle for understanding the security-level of a system is to analyze what information of importance can revealed via information-flow. Opacity is one of the most widely adopted information-flow security properties for dynamic systems whose information-flow is available to eavesdroppers or passive observers that are potentially malicious. Essentially, opacity captures the plausible deniability of the system for its   ``secret'' behavior by requiring that  the secret behavior can never be identified unambiguously by the intruder based on the online information-flow and the dynamic of the system. 

In this paper, we investigate  opacity in the context of discrete event systems (DES), which has drawn considerable attentions in the last decade; see, e.g., the recent surveys \cite{jacob2016overview,lafortune2018history} and the textbook \cite{hadjicostis2020estimation}.  
The basic concept of opacity was originally introduced by \cite{bryans2005modelling,bryans2008opacity} for dynamic systems modeled as transition systems. 
Depending on different secret requirements, different notions of opacity were proposed in the literature, including, e.g.,  initial-state opacity \cite{saboori2013verification,cong2019line}, current-state opacity \cite{wu2013comparative} and  $K$/infinite-step opacity \cite{saboori2011verification,saboori2012verification,yin2017new}.  
For example, current-state opacity requires that the outsider/intruder can never determine for sure that the system is at a secret state based on the information-flow. 
When the originally system is not opaque, many different approaches  have also  been proposed for the enforcement of opacity; see, e.g., \cite{takai2008formula,dubreil2010supervisory,falcone2015enforcement,barcelos2018enforcing,ji2018enforcement,liu2020k,zinck2020enforcing,mohajerani2020compositional}.  
More recently, the concept of opacity has been further generalized to continuous dynamic systems with infinite state-spaces and time-driven dynamics; see, e.g.,  \cite{ramasubramanian2020notions,an2020opacity,yin2020approximate,liu2020verification}.
In this work, we will study the verification of current-state opacity for finite systems. 

Since the essence of opacity is an information-flow security property, the underlying observation model of the system becomes the key in its modeling and analysis. In the DES literature, there are three different types of observation models that have been investigated:  static observation,  dynamic observation and  Orwellian observation. 
These three types were originally categorized in \cite{bryans2008opacity}. Here we summarize and explain them in detail as follows. 
\begin{itemize}
	\item 
	Static Observation:  
	In this setting, it is assumed that the event set is partitioned as observable and unobservable events, and the outsider can observe  observable events immediately upon their occurrences.  
	Such an observation model is called static because whether or not an event can be observed is fixed. Hence, the information-flow of a generated internal string is essentially its natural projection.  This is the most simple but probably also the most widely investigated observation model for the analysis of opacity; see, e.g., \cite{lin2011opacity,chen2017quantification,tong2017verification,keroglou2016probabilistic,yin2019infinite,balun2019opacity,saadaoui2020current,mohajerani2020ransforming}.  \vspace{8pt}
	\item 
	Dynamic Observation: 
	In this setting, it is assumed that whether or not the occurrence of an event can be observed is not fixed a priori and may depend on the \emph{prefix} string up to the current instant. 
	This model can describe the scenario where the observability of each event is  ``controlled" by an active information acquisition module. 
	Such an information acquisition module is also referred to as the sensor activation policy \cite{thorsley2007active,wang2010optimal,yin2019general,lin2020information},  the dynamic mask \cite{cassez2012synthesis,yin2020synthesis}  or the information release module \cite{zhang2015max,behinaein2019optimal} in the literature depending on the context of the underlying problem.  
	\vspace{8pt}
	\item 
	Orwellian Observation: Compared with the dynamic observation, in the setting of Orwellian observation, the observability of an event not only depends on the prefix up to the current instant, but also depends on the future \emph{suffix}. In other words, the system is allowed to re-interpret the observation of an internal string. In general, however, deciding opacity under an arbitrary Orwellian observation is undecidable even for finite systems \cite{bryans2008opacity}. In \cite{mullins2014opacity,yeddes2016enforcing}, the authors formulated a simple  Orwellian-type observation  using the notion of downgrading events, and showed that the corresponding opacity verification problem is decidable by  relating it to the notion of  intransitive non-interference \cite{hadj2005characterizing,hadj2005verification}. In \cite{berard2014verification}, the notion of rational observation is proposed, which  further generalizes the Orwellian projection proposed  in  \cite{mullins2014opacity}.
\end{itemize}

In this paper, we are interested in the verification of current-state opacity under  Orwellian-type observation.    
As we discussed above, the main feature of  Orwellian observation is that it allows to re-interpret the information of a previous event. 
Therefore, it is extremely useful in the modeling and analysis of information-flow security related to \emph{information declassification}. 
For example, the user can  classify the  occurrence of an event and declassify/release it in the future when some particular conditions are fulfilled.  
This is a very common way how secure information is processed.   
However, existing Orwellian-type observations as well as their associated definitions of opacity cannot fully capture   this scenario. 
For example, in the definition of  Orwellian projection  in  \cite{mullins2014opacity},  those ``on hold" information are released when a downgrading event occurs.  
However, in a more general setting, when to release those ``on hold" information  may be determined by a ``controller" having its own logic rather than simply depending on the event of the original system. Furthermore, in  \cite{mullins2014opacity},  it is assumed that once a downgrading event occurs, the entire trajectory is revealed, i.e., the outsider knows the current state precisely.  In practice, however, it is possible that only partial history information is  declassified and the outsider may still have ambiguity after the declassification. 

More importantly, existing definitions of current-state opacity under Orwellian-type observations still follow the standard projection-based definition by requiring that 
for any secret string, there exists a non-secret string such that they have the \emph{same projection}, e.g., \cite{mullins2014opacity,berard2014verification}. 
However, this definition is not exactly suitable for Orwellian-type observation since the observation sequence may not be prefix-closed. 
In particular,  that two strings have the same projection does not necessarily imply that their prefixes also have the same projection. If the prefixes of two projection-equivalent strings are not projection-equivalent, then the intruder may still be able to determine that the system is currently at a secret state even when the standard projection-based opacity condition holds.  
This scenario will be explained in more detail in Section~\ref{sec3}. 
Therefore, new definition of opacity as well as the associated verification algorithms are needed for  Orwellian-type observation.  

Motivated by the above discussions, in this paper, we propose a new Orwellian-type observation framework for information-flow security and investigate the underlying opacity verification problem. Compared with the existing works, the main contributions of this work are as follows.
\begin{itemize}
	\item 
	First, we propose a new Orwellian-type observation model called  the \emph{Dynamic Information Release Mechanism} (DIRM).   
	The DIRM is motivated by \cite{mullins2014opacity} but has the following main difference. 
	Compared with  \cite{mullins2014opacity} where downgrading events are used to trigger information release, 
	in the DIRM, when to release those ``hold on" information is \emph{state-dependent}. 
	This model captures the scenario where information declassification can be controlled by another agent that is not embedded in the original plant model.  
	Furthermore, we investigate the partial information release setting, 
	which is more general than the full information release setting in \cite{mullins2014opacity}, 
	where entire trajectory will be available when the information is released.
	\vspace{8pt}
	\item 
	Second, we propose a new definition of current-state opacity that is more suitable for Orwellian-type observations.  
	Specifically, we introduce  the concept of  observation history as the set of all projections along the prefixes of a string.  
	Then we say that a system is current-state opaque if for any secret string, there exists a non-secret string such that they have the same \emph{observation history}.  
	The new  history-based definition is more suitable than the standard projection-based definition, since the observation history may not be prefix-closed for Orwellian-type observations.   
    \vspace{8pt}
	\item 
	Finally, we propose an effective algorithm for the verification of the proposed notion of current-state opacity under the DIRM. 
	Our approach is based on the notion of augmented system  and   a new information structure called the DIRM-observer.  
	Specifically, the augmented system augments the original state-space by an additional binary information that tracks whether or not there is information to be released on hold. 
	The DIRM-observer essentially tracks both the current-state estimate and the information that is held on. 
	When an information-release-state is reached, it updates the current-state-estimate by effectively fusing these two parts of information together. 
	Then we show that the opacity verification problem can be solved by a simply reachability search within the DIRM-observer. 
\end{itemize}

The rest of this paper is organized as follows.
In Section~\ref{sec2},  some necessary preliminaries are introduced. 
In Section~\ref{sec3}, we present  the dynamic information release mechanism as the underlying Orwellian-type observation model and define current-state opacity under the DIRM projection. 
To verify opacity, we first introduce the concept of augment system in Section~\ref{sec4} and show that the verification problem can be investigated equivalently for the augmented system. 
In Section~\ref{sec5}, we present the  DIRM-observer  structure and show how to use it to verify opacity.
A case study on security issues in  federated cloud computing systems is provided in Section~\ref{sec6}. 
Finally, we conclude the paper in Section~\ref{sec7}.

\section{Preliminaries}\label{sec2}

\subsection{System Model}

Let $\Sigma$ be a finite set of events. A string is a finite sequence of events and we denote by $\Sigma^*$ the set of all strings over $\Sigma$ including the empty string $\epsilon$.  
For any string $s=\sigma_1\sigma_2\cdots \sigma_n$, 
we denote by $ |s| $  the length of string $ s $, i.e., $|s|=n$ and $|\epsilon|=0$; 
we also denote by  
$s[i,j]$ the sub-string of $s$ from the $i$th event to the $j$th event, i.e., 
$s[i,j]=\sigma_i\sigma_{i+1}\cdots\sigma_j$. 
In particular, we define $s[i,j]=\epsilon$ if $j<i$ 
and define $s[i,j]=s[i,|s|]$ if $|s|<j$.  
A language $L\subseteq \Sigma^*$ is a set of strings. 
For any language $ L \subseteq \Sigma^* $, we denote by $\overline{L}$ its  \textit{prefix-closure}, i.e., $ \overline{L}=\{t\in \Sigma^*:\exists w\in \Sigma^*$ s.t. $ tw\in L \} $.

We consider a DES modeled as a deterministic finite-state automaton (DFA)
\[
G=(X,\Sigma,\delta,x_0), 
\]
where $ X $ is a finite set of states, $\Sigma$ is a finite set of events, $\delta: X \times \Sigma \to X$ is the partial transition function, and $x_0\in X$ is the initial state.
For any $ x , x' $ $\in X$, $\sigma \in \Sigma$, $\delta(x,\sigma)=x'$  means that there exists  a transition from state $ x $ to state $x'$ labeled with  event $\sigma$.
For the sake of simplicity, we write $ {\delta}( x,s) $ as $ {\delta}(s) $ when $x={x}_0$.
The transition function is also extended to $\delta:X\times\Sigma^*\to X$ recursively by: 
$\delta(x,\epsilon)=x$ and  $\delta(x,s\sigma)=\delta(\delta(x,s),\sigma)$. 
The language generated by $G$ is $\mathcal{L}(G) = \{s \in \Sigma^{*} : \delta(x_0,s)!\}$, where ``$!$" means ``is defined".

\subsection{Current-State Opacity under Natural Projection}
In the analysis of information-flow security, it is often assumed that the event set is partitioned as 
$ \Sigma=\Sigma_o\dot{\cup}\Sigma_{uo} $, 
where $\Sigma_{o}$ is the set of observable events and  $\Sigma_{uo}$ is the set of unobservable events.  
Let $\hat{\Sigma}\subseteq\Sigma$ be a set of events. 
Then the natural projection from $\Sigma$ to $\hat{\Sigma}$ is a mapping $P_{\hat{\Sigma}}: \Sigma^{*}\to \hat{\Sigma}^{*}$ defined recursively by: for any $s\in \Sigma^*$, $\sigma\in \Sigma$, we have 
\begin{align*}
P_{\hat{\Sigma}}(\epsilon)=\epsilon \text{ \ and \ }
P_{\hat{\Sigma}}(s\sigma)=
\left\{
\begin{array}{l l}
P_{\hat{\Sigma}}(s)\sigma\quad
&\text{if } \sigma \in   \hat{\Sigma}   \\
P_{\hat{\Sigma}}(s)\quad
&\text{if } \sigma \notin  \hat{\Sigma}
\end{array}
\right..
\end{align*}
In the context of opacity, we assume the underlying system has some ``secret"  modeled as a set of secret states $ X_{S} \subset X $. 
Furthermore, there is a passive intruder modeled as an observer that knows the system model and can eavesdrop  the projected information flow, i.e., it can observe the occurrences of events in $\Sigma_o$.  
Then current-state opacity requires that the intruder should never know for sure that the system is currently at a secret state, which is defined as follows.
\begin{definition}\label{def:opa-standard}\upshape 
	Given system $ G $ and  secret states $  X_S \subset X$, 
	we say system $ G $ is  \emph{current-state opaque} (w.r.t. $ X_S $ and $\Sigma_o$), if for any string $s\in \mathcal{L}(G)$ such that $\delta(s)\in X_S$,  there exists another string $s'\in \mathcal{L}(G)$ such that $\delta(s')\in X\setminus X_S$  and  $P_{\Sigma_o} (s)=P_{\Sigma_o} (s')$. 
\end{definition}

\section{Opacity under Dynamic Information Release Mechanism}\label{sec3}

\subsection{Dynamic Information Release Mechanism} 
In the standard analysis of opacity under natural projection, the observation of each observable event is assumed to be \emph{instant} in the sense that the occurrence of each observable event can be observed immediately and the occurrence of each unobservable event can never be observed. 
In some applications, however, the event set cannot be simply partitioned as observable and unobservable.    
In practice, the occurrence of some event may be  classified until the system decides to \emph{release (or declassify)} it. 
This is a very common scenario in information-flow security problem, e.g., some secure documents will be declassified only when certain conditions are satisfied. 
This leads to the  \emph{dynamic information release mechanism} (DIRM) defined as follows. 

Formally, we assume that the event set is partitioned as 
\[
\Sigma=\Sigma_o\dot{\cup}\Sigma_{uo}\dot{\cup}\Sigma_{r}, 
\]
where
\begin{itemize}
	\item 
	$\Sigma_o$ is the standard set of observable events whose occurrences can be observed instantly; 
	\item 
	$\Sigma_{uo}$ is the standard set of unobservable events whose occurrences can never be observed;
	\item 
	$\Sigma_r$ is the set of events whose occurrences may not be observed  instantly but can be \emph{released} in the future.  	 
\end{itemize}
To formally describe how the information is released, we consider a \emph{state-based} dynamic information release mechanism specified by a function
\[
R:X \to \{\textbf{u},\textbf{r}  \}, 
\]
where  ``\textbf{r}" and ``\textbf{u}" stand for ``release" and ``unrelease", respectively. 
One can image that the system has an ``information release button" for events in $\Sigma_r$ and the release button is pressed when a state $x$ such that $R(x)=\textbf{r}$ is encountered. In such a case, all events in $\Sigma_r$ along the trajectory that have not yet be observed will be released   at state $x$ in the sense that the observer knows when they occurred.

\begin{remark}	\upshape
	Here we use function $R$   to emphasize the fact that the information release mechanism is \emph{state dependent}. Equivalently, we can also define  $X_r=\{x\in X: R(x)=\textbf{r} \}$ as the set of states at which $\Sigma_r$ can be released. Hereafter, we will only use $X_r$ instead of function $R$ and also refer to $X_r$ as the DIRM, for the sake of simplicity.  
	Also, without loss of generality but for the sake of simplicity, 
	an event will not be released immediately after its occurrence, i.e., $\forall s\in \mathcal{L}(G),\sigma \in \Sigma_r$: $\delta(s,\sigma)\notin  X_r$.   
\end{remark}

\begin{remark}\upshape
	The dynamic information release mechanism  defined here is state-based. 
	In general, it can be a language-based function specified by a finite-state transducer. In this case, one can take the product of the transducer with the plant to refine the state-space such that the release mechanism becomes state-based. 	
\end{remark}

\subsection{The DIRM Projection}
Let $s  \in \mathcal{L}(G) $ be a string generated by the system.  
We denote by $0\leq \imath_s\leq |s|$ the latest instant  when events in $\Sigma_r$ are released, i.e., \vspace{-5pt}
\begin{itemize}
	\item 
	$\delta(x_0,   s[1,\imath_s] )  \in  X_r$; and \vspace{5pt}
	\item 
	$\forall \imath_s< i\leq |s| :  \delta(x_0,   s[1, i  ]  )  \notin  X_r$. 
\end{itemize} 

In the DIRM, the observation of a string $s$ depends on its release status. 
We define the corresponding DIRM projection $P_R(s)$ of it as follows.

\begin{definition}\label{def2}\upshape  
	Given system $G$ with partition $\Sigma=\Sigma_o\dot{\cup}\Sigma_{uo}\dot{\cup}\Sigma_{r}$
	and DIRM $X_r\subseteq X$, 
	the \textit{DIRM projection} of string $s$, denote by $P_R(s)$, is defined by 
	\begin{equation}
	P_R(s)=  P_{\Sigma_o\cup \Sigma_r}( s[1, \imath_{s}]   )P_{\Sigma_o}( s[\imath_{s}+1,|s|]   )
	\end{equation} 		
\end{definition} 
The intuition of the above definition is as follows. 
Recall that $\imath_{s}$ is the latest information release instant when $s$ is executed. Therefore, events in $\Sigma_r$ that occur before instant $\imath_{s}$ are released, 
which is captured by $P_{\Sigma_o\cup \Sigma_r}( s[1, \imath_{s}]   )$,  while events in $\Sigma_r$ that occur after  $\imath_{s}$ are still unobservable, which is captured by $P_{\Sigma_o}( s[\imath_{s}+1,|s|]   )$. 
Note that, when $\imath_{s}= |s|$, by definition we have  $ s[1, \imath_{s}]=s$ and 
$s[\imath_{s}+1,|s|]= \epsilon$. Then in this case, we have $P_R(s)= P_{\Sigma_o\cup \Sigma_r}(s)$ as all events in $\Sigma_r$ are released. 
Also, when $\imath_{s}=0$, it means that there is no information released along the path. 
Therefore, by definition, we have $s[1,\imath_{s}]=\epsilon$ and $s[\imath_{s}+1,|s|]=s$, which means that $P_R(s)=P_{\Sigma_o}(s)$. 
Also, the above definition implicitly assumes the DIRM has infinite memory for those unreleased information. 
We illustrate the DIRM projection by the following example.

\begin{figure} 
	\centering
	\includegraphics[width=0.37\textwidth]{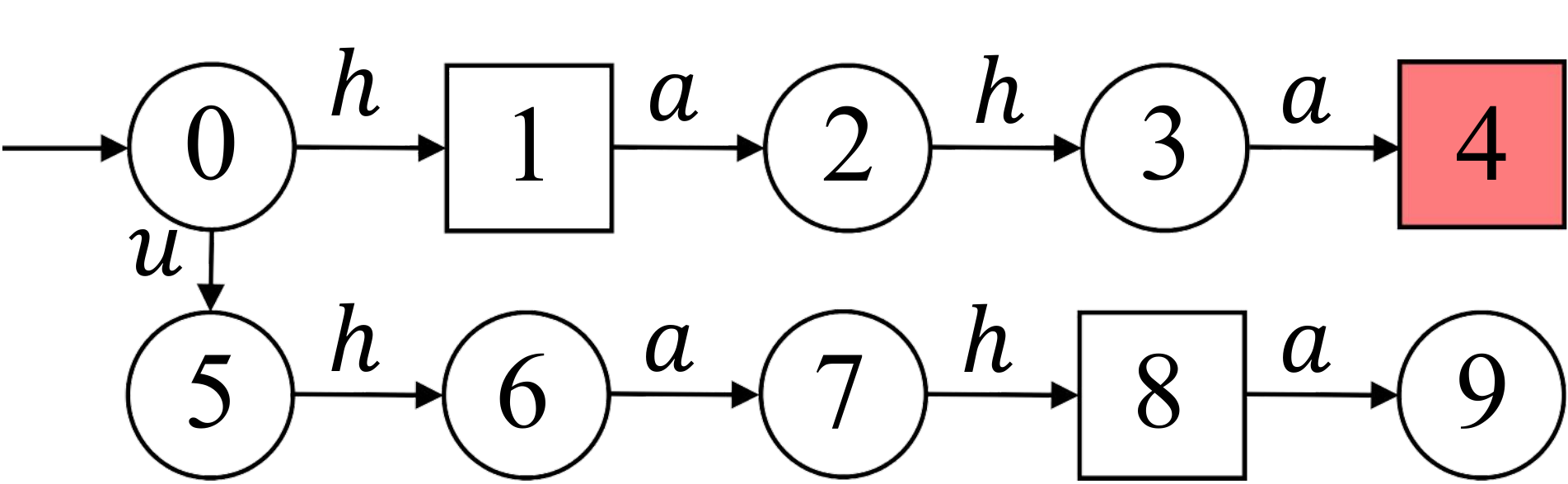} 
	\caption{System $ G $ with $\Sigma_{o}=\{a\}, \Sigma_{r}=\{h\}$, $\Sigma_{uo}=\{u\}$, $X_r=\{1,4,8\}$ and $ X_S=\{4\} $.}  
	\label{ex1}  
\end{figure}

\begin{example}\label{example:2}\upshape
	Let us consider system $ G $ shown in Figure~\ref{ex1}, 
	where $ X_r = \{1,4,8\} $, $\Sigma_{o}=\{a\}$, $\Sigma_{r}=\{h\}$, $\Sigma_{uo}=\{u\}$  and $ X_S=\{4\} $.  
	We use rectangles to denote states in $X_r$. 
	For string $ s=uh $, we have $ \imath_{s}=0 $ since $\forall s'\in \overline{\{s\}}:   \delta(x_0,  s' )\notin X_r$, i.e., no state in $X_r$ is visited along $s$ from the initial state. 
	Therefore, we have 
	\[
	P_R(uh)= P_{\Sigma_o\cup \Sigma_r}( s[1, \imath_{s}]   )P_{\Sigma_o}( s[\imath_{s}+1,|s|]   )=
	P_{\Sigma_o}( uh  )=\epsilon, 
	\]
	i.e., no event is observed upon the occurrence of $uh$. 
	
	Comparatively, for string $ s= h $, we have $ \imath_{s}=1 $ since string $h$ reaches state $1\in X_r$. 
	Then we have
	\[
	P_R(h)= P_{\Sigma_o\cup \Sigma_r}( s[1, \imath_{s}]   )P_{\Sigma_o}( s[\imath_{s}+1,|s|]   )=
	P_{\Sigma_o\cup \Sigma_r}(h)=h. 
	\]
	This is because the occurrence  of $h$ is released when state $1\in X_r$ is reached. 
	Note that events in $\Sigma_{r}$ that occur after reaching $ X_r $ will become unobserverable again until a new state in $ X_r $ is reached again.  For example,  for string $ s=hah $, we have still have  $ \imath_{s}=1 $, which gives
	\begin{align}
	P_R(hah )= &P_{\Sigma_o\cup \Sigma_r}( s[1, \imath_{s}]   )P_{\Sigma_o}( s[\imath_{s}+1,|s|]   )\nonumber\\
	=& P_{\Sigma_o\cup \Sigma_r}( s[1, 1]   )P_{\Sigma_o}( s[2,3] )\nonumber\\
	=& P_{\Sigma_o\cup \Sigma_r}(h) P_{\Sigma_o}(ah)\nonumber\\
	=&ha\nonumber
	\end{align}
	That is, only the first occurrence of $h$ is released, while the second is not.  
	\hfill $\qed$
\end{example}

\begin{remark}\upshape
According to the definition of the DIRM projection, 
once an event in $\Sigma_r$ is released, the outside observer not only knows the occurrence of this event, but also knows \emph{when} it occurred. 
This is different from the case of   delayed observations, where  events are 
``held on" physically, e.g., in  communication channels, 
and the released events are observed following the existing observation. 
However, in our DIRM,  the current observation can be overwritten after information release because we know when these previous events occurred.    
\end{remark} 


\subsection{Opacity under DIRM Projection}
In order to extend current-state opacity from the standard natural projection to  our DIRM projection, one may think it suffices to replace $P_{\Sigma_o}$ in Definition~\ref{def:opa-standard} by $P_{R}$. However, the following example shows that such a definition is not  suitable for the case of dynamic information release mechanism.
\begin{example}\upshape  Again, let us still consider system $ G $ shown in Figure~\ref{ex1}.
	For string $ s=haha $, there exists another string $ s'=uhaha $ such that $ P_R(s)=P_R(s')=haha $. 
	If we simply replace   $P_{\Sigma_o}$ in Definition~\ref{def:opa-standard} by $P_{R}$, then we will assert that the system is current-state opaque
	because for string $ s=haha $, which is the only string that leads to secret state $4$, there exists another string $ s'=uhaha $ that leads to a  non-secret and they have the same projection.  
	
	However, this definition is not suitable here as the intruder can still assert that the system is currently at secret state $4$ when string $ s=haha $ is executed.  To see this point more clearly, let us consider what the intruder can observe when string $ s=haha $ is generated. Initially, it observes $h$ when string $h$, which reaches $1\in X_r$, is executed. Then it observes $ha$ when $a$ occurs. Finally, the intruder observes $haha$ when state $4\in X_r$ is reached.  
	On the other hand, for string $ s'=uhaha $, the intruder will not observe $h$ when $uh$ is executed since this string leads to $6\notin X_r$. 
	Therefore, the intruder will first observe $a$ when string $uha$ is executed. When the second $h$ is executed, which leads to state $8\in X_r$, the intruder observes $hah$, i.e., the first occurrence of $h$ was ``held on" until this point. 
	Therefore, the information histories available to the intruder are different when strings $s$ and $s'$ are executed. 
	In other words, once the intruder observes $h$ directly, it knows for sure that the system is currently at state $1$. 
	Therefore, if it further observes $haha$, it can still determine that the system is at secret state $4$, which makes the system non-opaque. 
	\hfill $\qed$
\end{example}

The above example reveals an important feature in the DIRM projection. 
That is, 
for any two strings $s,s'\in \mathcal{L}(G)$,  $P_R(s)=P_R(s')$ does not necessarily imply that $s$ and $s'$ generates the same information-flow. 
This is because they may have different \emph{intermediate observations} and hence, these two strings can still be distinguished. 
Note that such an issue does not arise in the standard natural projection as two strings having the same projection must have the same projection for their prefixes with the same length. 
However, it is not the case for our DIRM projection, which is the key technical issue we need to handle. 

To correctly capture the information-flow of a string $s \in \mathcal{L}(G)$, we define
\[
H_R(s)=\{ P_R(s') : s'\in \overline{\{s\}} \}
\]
as the \emph{history} of string $s$. 
For the case of natural projection, 
history equivalence and projection equivalence are the same.
However,  in our setting, history equivalence is stronger than projection equivalence as it requires that the projections of all prefixes, i.e., all intermediate observations, are also equivalent. 

\begin{remark}\upshape
	Here, a history is defined as an unordered set rather than an order sequence according to the ordering of each observation. 
	This definition is without loss of generality as the length of $P_R(s')$ is always non-decreasing when the length of the prefix $s'$ increases. Therefore, one can always recover the ordering of the observation, if needed,  based on the length of each element in the history set $H_R(s)$.  
\end{remark}

Therefore, to define current-state opacity, one can replace $P_{\Sigma_o}(\cdot)$ in Definition~\ref{def:opa-standard} by $H_R(\cdot)$. Here, we provide an equivalent definition using the notion of  current-state estimate (CSE)  for the sake of future developments. 
Formally, for any string $s\in \mathcal{L}(G)$, the current-state estimate upon  history   $H_R(s)$ is defined by 
\[ 
\hat{X}(H_R(s)) =	\{\delta(s') \!\in\! X: \exists s'\!\in \!\mathcal{L}(G)\text{ s.t. }   H_R(s)\!=\!H_R(s')\}.
\]

Then we introduce current-state opacity under DIRM using the CSE as follows.
\begin{definition}\label{def:opa-DIRM} \upshape
	Given   system $ G $ with  secret states $  X_S \subset X$ and DIRM $X_r \subseteq X$, 
	we say system $ G $ is  \emph{current-state opaque} (w.r.t. $ X_S$ and DIRM $X_r$), if 
	\[
	\forall s\in \mathcal{L}(G):  \hat{X}(H_R(s)) \nsubseteq X_S .
	\] 
\end{definition}
According to the definition of opacity,  we implicitly assume that the intruder is aware of the DIRM  of the underlying system.  
We illustrate this definition by the following example. 
\begin{example}\upshape
	Let us still consider system $ G $ in Figure \ref{ex1}. 
	Then for string $ s=haha$ leading to  secret state $4$, 
	we have $ H_R(s)=\{\epsilon, h,ha,haha \} $. 
	Note that $s$ is the only string that generates   history $H_R(s)$, e.g.,  
	for string $ s'=uhaha$, we have  $H_R(s')=\{\epsilon, a, hah, haha\}\neq H_R(s)$. 
	Therefore, we have   $\hat{X}(H_R(s))=\{4\}\subseteq X_S$, which means that system $ G $ is not current-state opaque with respect to DIRM $X_r$ and secret $X_S$.

	Comparatively, suppose that $ X_S=\{2\}$, then the system becomes opaque. 
	This is because  $ t=ha $  is the only string that leads to secret state $2$. 
	However, we also have $t'=hah$ such that  $H_R(t)=H_R(t') =\{\epsilon, h, ha \}$ and $\delta(t')=3$. 
	Therefore, we have   $\hat{X}(H_R(t))=\{2,3\}\nsubseteq X_S$.
	\hfill $\qed$
\end{example}

\begin{remark}\upshape 
	Our DIRM projection is closely related to the so called \emph{Orwellian projection} proposed in \cite{mullins2014opacity}, 
	where a \emph{downgrading event} is used to trigger information release. 
	Our DIRM projection is different from the Orwellian projection in the following aspects. 
	First, we consider partial information release while the Orwellian projection considers full information release. 
	Specifically,  under the Orwellian projection, once the information release mechanism is triggered, the outsider knows immediately the precise state of the system. This is not the case of our DIRM projection since we assume that events in $\Sigma_{uo}$ are always unobservable, i.e., even when information release is triggered, the intruder may still have information uncertainty. 
	Also, we consider a \emph{state-based} information release mechanism rather than using a downgrading event to trigger the release as the case of \cite{mullins2014opacity}. We believe such a modeling is more natural in practice since, e.g.,  when to release information can be controlled by another transducer that does not need to be associated with events in the original plant model. Furthermore, the state-based mechanism also brings new technical challenges. 
	In particular, due to the use of downgrading event, two strings having the same projection necessarily have the same history. 
	However, for the general case of Orwellian-type observation,   we need to consider  histories rather than projections, which is the most significant technical challenge in our setting.
\end{remark}

Note that, in the standard natural projection setting, the current-state estimate can be computed easily by constructing the standard \emph{observer automaton}   that essentially recursively updates the belief of the observer on-the-fly; see, e.g., \cite{Lbook}. 
However, the computation of  $\hat{X}(H_R(s))$, which is the key to the opacity verification problem, is much more challenging as delayed information is involved. 
In the following sections, we will elaborate on how to compute  $\hat{X}(H_R(s))$.

\section{Augmenting the System with the Trajectory Information}\label{sec4}

\subsection{Augmented Systems}
We note that, not every visit of an information-release-state in $X_r$ will release some previous information; 
it depends on whether or not the trajectory visiting $X_r$ contains events in $\Sigma_r$ since the last visit of $X_r$, i.e., whether or not there are  events in $\Sigma_r$ that are ``on hold". 
To capture the ``true" information-release-states, we augment the original plant by an additional  binary information  $\{N,Y\}$ to obtain a new DFA  
\[
\tilde{G} = (\tilde{X},\Sigma,\tilde{\delta},\tilde{x}_0),
\] 
where
\begin{itemize}
	\item 
	$ \tilde{X}\subseteq X \times \{N,Y\} $ is the set of augmented states; 
	\item 
	$ \tilde{x}_0=(x_0,N) $ is the  initial augmented state;
	\item 
	$ \tilde{\delta}: \tilde{X} \times \Sigma \to \tilde{X} $ is the augmented transition function defined by:
	for any $\tilde{x} = (x,l)\in \tilde{X} $ and $\sigma \in \Sigma $ such that $\delta(x,\sigma)!$, we have   
	\[
	\tilde{\delta}(\tilde{x},\sigma)\!=\!
	\left\{  
	\begin{array}{l l}
	(\delta(x,\sigma),Y)\quad
	&\text{if } [\sigma \!\in \!\Sigma_{r}]\vee [x\!\notin\! X_r  \wedge l\!=\!Y]    \\
	(\delta(x,\sigma), N) \quad
	&\text{otherwise} 
	\end{array}
	\right.
	\]	
\end{itemize}
Intuitively, a state is augmented with labeled $Y$ if it is either visited directly via an event in $\Sigma_r$ or visited by a string that contains event $\Sigma_r$ since the last visit of information-release-states $X_r$. 
Therefore, a state augmented with $Y$ is a state at which some  historical events can be released. 
When the current state $x$ is in $X_r$ and the upcoming event $\sigma$ is not in $\Sigma_r$,  label $Y$ will be reset to $N$, which means that the previous information has been released (at state $x$) and there is no new information to be released in the future added. 
We illustrate the augmented system $\tilde{G}$ by the following example.

\begin{figure} 
	\centering
	\subfigure[System $G$]{\label{ex4-1}
		\includegraphics[width=0.17\textwidth]{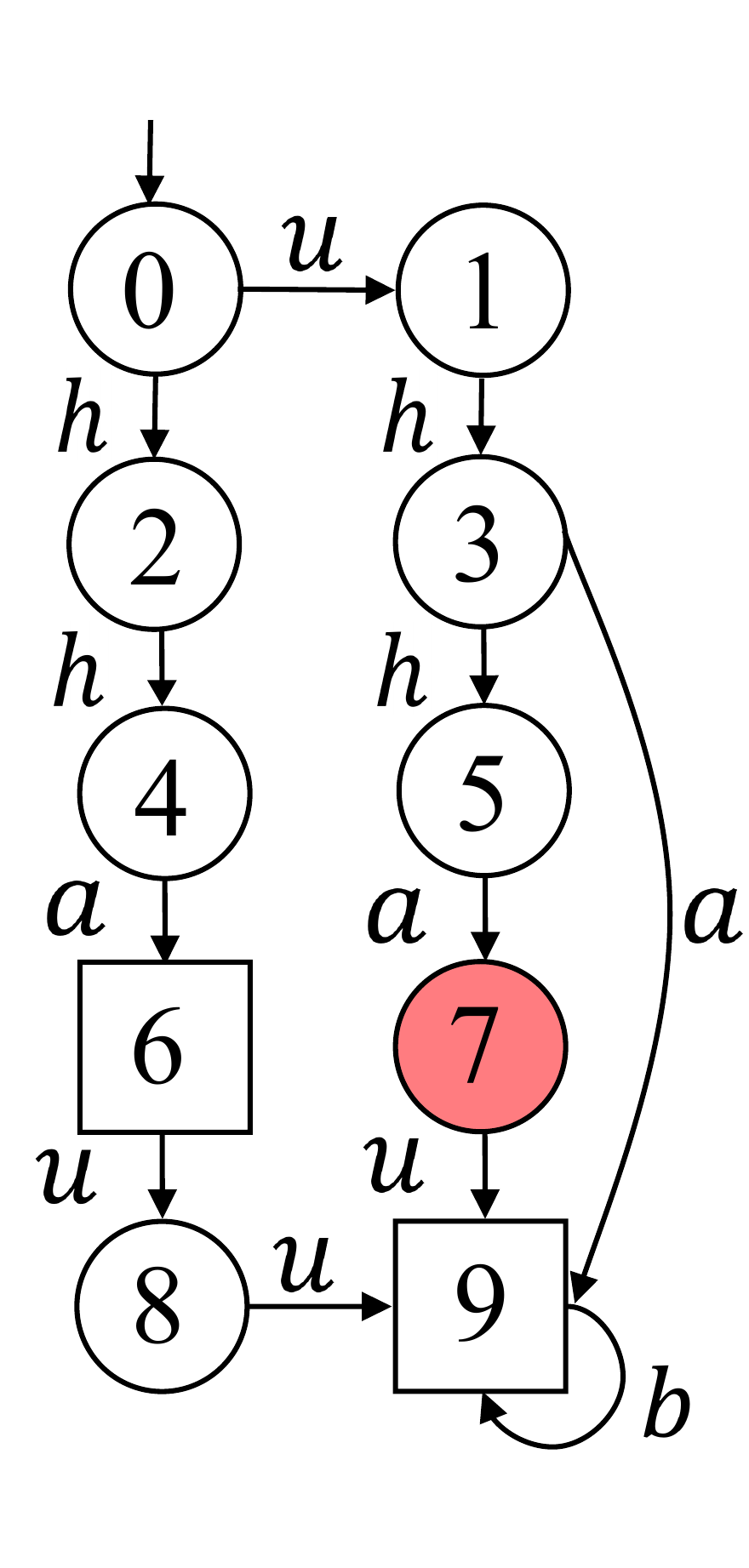}}
	\subfigure[Augmented system $\tilde{G}$]{\label{ex4-2}
		\includegraphics[width=0.22\textwidth]{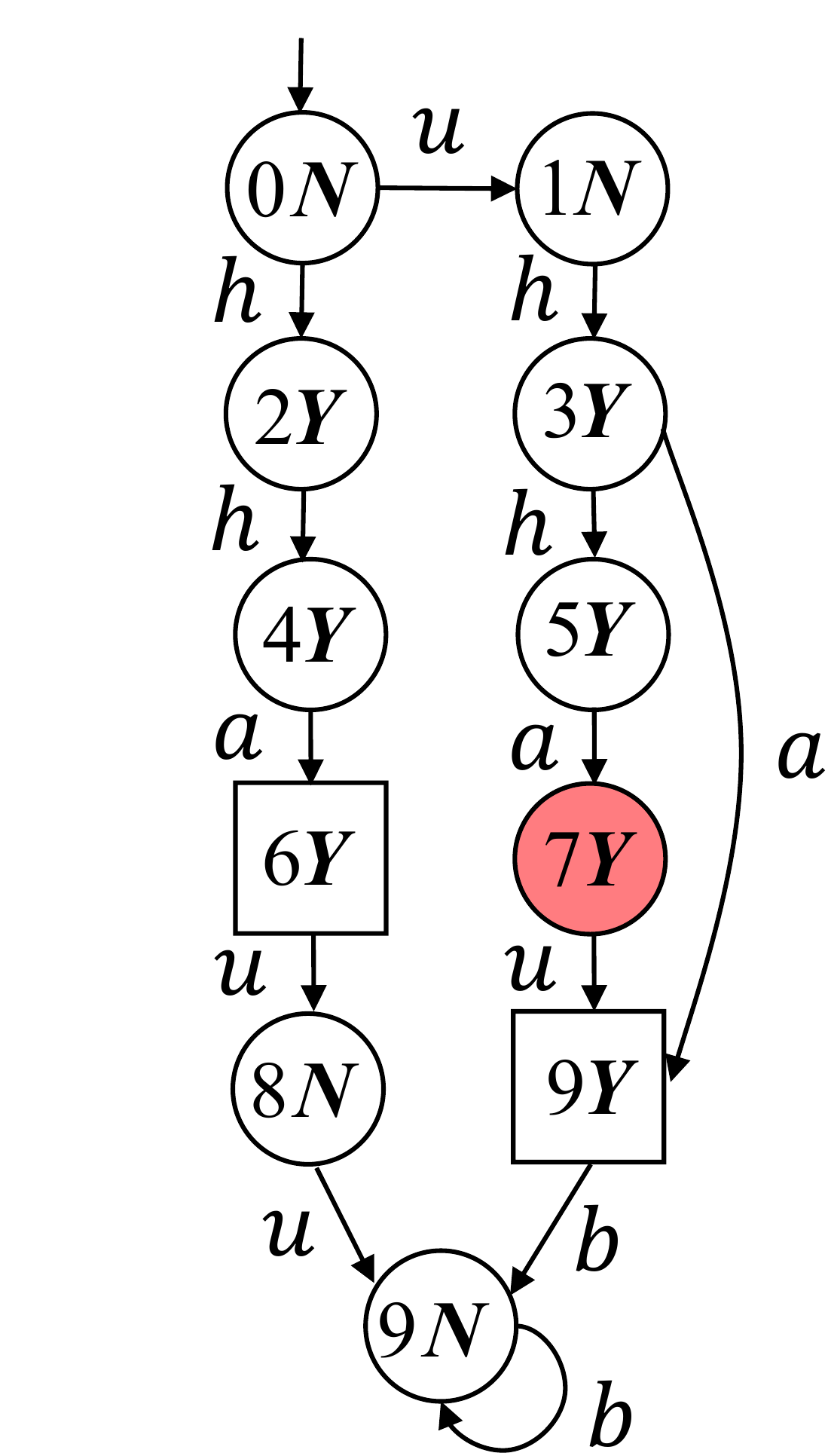}}
	\caption{System $ G $ and corresponding augmented system $\tilde{G}$, where $\Sigma_{o}=\{a,b\}$, $\Sigma_{r}=\{h\}$, and $\Sigma_{uo}=\{u\}$.}
	\label{ex4}
\end{figure}

\begin{example}\upshape 
	Let us consider DFA $ G =(X,\Sigma,\delta,x_0)$ shown in Figure~\ref{ex4-1}, 
	where   $\Sigma_{o}=\{a,b\}$, $\Sigma_{r}=\{h\}$, and $\Sigma_{uo}=\{u\}$.  
	The DIRM is specified by information-release-states $ X_r = \{6,9\} $. 
	Then the augmented system $\tilde{G}$ of $ G $ is depicted in Figure~\ref{ex4-2}.
	For the sake of simplicity, in the figure, we omit  brackets and comma for each state in $\tilde{X}$, e.g., state  $ (1,Y) $ is simplified as $ 1Y $. 
	
	Specifically, the initial state is $0N$ and when event $u$ occurs, we reach state $1N$. 
	However, when $r$ occurs, we reach state $2Y$ since $h\in \Sigma_r$. 
	From state $6Y$, when event $u$ occurs, we reach state $8N$ since $6\in X_r$ and $u\notin \Sigma_r$. 
	Note that state $9$ in $ G $ is split as $ 9N $ and $9Y$ in $\tilde{G}$ depending on how it is visited.
	\hfill $\qed$
\end{example}

The augmented system $\tilde{G}$ contains at most $2\cdot|X|$ states which is linear in the size of the original system. 
Furthermore, since for any state $(x,l)\in \tilde{X}$,   $\tilde{\delta}((x,l),\sigma)$ is defined if and only if  $\delta(x,\sigma)$ is defined, we also have 
\[\mathcal{L}(\tilde{G})=\mathcal{L}(G), \]
i.e., $\tilde{G}$ is essentially a state-space refinement of $G$.  
Hereafter, we will analyze opacity based on the augmented system $\tilde{G}$ instead of $G$. 
To this end, we define
\[
\tilde{X}_S=\{(x,l)\in \tilde{X}:  x\in X_S  \}
\]
as the set of secret states in $\tilde{G}$. 
Also, we define 
\[
\tilde{X}_r=\{(x,l)\in \tilde{X}:  x\in X_r \text{ and }l=Y \}
\]
as the set of information-release-states in $\tilde{G}$. 

To distinguish with notations for $ G $, before the end of this subsection, we first use $\tilde{P}_R(\cdot)$, $\tilde{H}_R(\cdot) $, $\hat{\tilde{X}}(\tilde{H}_R(s))$  to represent, respectively, as the DIRM projection, the history and the current-state estimate  w.r.t.\ $ \tilde{G} $ and $\tilde{X}_r$. 
The following result shows that considering the augmented system $\tilde{G}$ and the corresponding DIRM $\tilde{X}_r$ is the same as the original system.  

\begin{lemma}\label{lem1}\upshape
	For any string $s\in \mathcal{L}(G)= \mathcal{L}(\tilde{G})$, we have $H_R(s)=\tilde{H}_R(s)$.
\end{lemma}
\begin{pf}	
	By  definition, we have $ H_R(s)=\{ P_R(s') : s'\in \overline{\{s\}} \} $ and $ \tilde{H}_R(s)=\{ \tilde{P}_R(s') : s'\in \overline{\{s\}} \} $, where  for any $ s'\in \overline{\{s\}}$, we have $ P_R(s')= P_{\Sigma_o\cup \Sigma_r}( s'[1, \imath_{s'}]   )P_{\Sigma_o}( s'[\imath_{s'}+1,|s'|]   ) $ and $ \tilde{P}_R(s')=  P_{\Sigma_o\cup \Sigma_r}( s'[1, \tilde{\imath}_{s'}]   )P_{\Sigma_o}( s'[\tilde{\imath}_{s'}+1,|s'|]   ) $.
	Here we use  $ \imath_{s'} $ and $\tilde{\imath}_{s'} $ to denote the latest instant  when $s'$ reaches $X_r$ in $G$ and the latest instant when $s'$ reaches $\tilde{X}_r$ in $\tilde{G}$, respectively. 
	Since $\tilde{X}_r \subseteq X_r \times \{Y\} $, we have $ \imath_{s'} \geq \tilde{\imath}_{s'} $. 
	
	To show that $H_R(s)=\tilde{H}_R(s)$, it suffices to show that   $P_R(s')=\tilde{P}_R(s')$ for any $s'\in \overline{\{s\}}$. 
	Note that, if $ \tilde{\imath}_{s'} = \imath_{s'} $, we have immediately that  $ P_R(s')=\tilde{P}_R(s') $. Therefore, we consider the case of  $ \imath_{s'} > \tilde{\imath}_{s'} $ hereafter. 
	For this case, we can write $P_R(s')$ and $\tilde{P}_R(s')$  in the forms of
	\begin{align}
	P_R(s')&=  
	P_{\Sigma_o\cup \Sigma_r}( s'_1 )P_{\Sigma_o\cup \Sigma_r}( s'_2) P_{\Sigma_o}(s'_3)\nonumber\\
	\tilde{P}_R(s')&=   
	P_{\Sigma_o\cup \Sigma_r}( s'_1 ) 	P_{\Sigma_o}( s'_2) P_{\Sigma_o}( s'_3),\nonumber
	\end{align}
	where $s'_1=s'[1, \tilde{\imath}_{s'}], s'_2=s'[\tilde{\imath}_{s'}+1,\imath_{s'}]$ and $s'_3= s'[\imath_{s'}+1,|s'|]$.
	Therefore, it remains to show that 
	$ P_{\Sigma_o}( s'_2)=P_{\Sigma_o\cup \Sigma_r}(s'_2) $, i.e., 
	$s'_2$	does not contain events in $\Sigma_r$. 
	
	Let $s_2'=\sigma_1\dots\sigma_m$ and let $x_i=\tilde{\delta}(s_1'\sigma_1\cdots\sigma_i)$ be the $i$th state visited from $\tilde{\delta}(s_1')\in \tilde{X}_r$ in $\tilde{G}$. 
	Assume for the sake of contradiction that $s_2'$ contains an event in $\Sigma_r$ and let $k\in \{1,\dots,m\}$ be the largest number such that 
	$\sigma_k\in \Sigma_r$. 
	Furthermore, let $p\in \{k,\dots,m\}$ be the smallest instant such that
	$x_p\in X_r$. 
	Note that $x_p$ is well-defined since $x_m=\tilde{\delta}(s_1's_2')\in X_r$. 
	Then, by the definition of $\tilde{G}$, we have 
	$x_p \in X_r\times \{Y\}=\tilde{X}_r$. 
	However, this violates the fact that $\tilde{\imath}_{s'} $ is the latest instant when $s'$   reaches $\tilde{X}_r$ in $\tilde{G}$. 
	Therefore, $s_2'$ does not contain any event in $\Sigma_r$, which completes the proof.\hfill $\qed$
\end{pf}

Then the following theorem shows that to check opacity for $G$, 
it suffices to check opacity for $\tilde{G}$. 

\begin{theorem}\label{theo1}
	$G$ is  current-state opaque w.r.t.\ secret states $X_S$ and DIRM $X_r$, if and only if, 
	$\tilde{G}$ is  current-state opaque w.r.t.\ secret states $\tilde{X}_S$ and DIRM $\tilde{X}_r$.
\end{theorem}
\begin{pf}
	For any string $s\in \mathcal{L}(G)=\mathcal{L}(\tilde{G})$, the current-state estimate of $G$  by observing $H_R(s)$ is 
	\[
	\hat{X}(H_R(s)) =	\{\delta(s') \!\in\! X: \exists s'\!\in \!\mathcal{L}(G)\text{ s.t. }   H_R(s)\!=\!H_R(s')\}
	\]	
	and the current-state estimate of $\tilde{G}$  by observing $\tilde{H}_R(s)$ is	
	\[
	\hat{\tilde{X}}(\tilde{H}_R(s)) =	\{\tilde{\delta}(s') \!\in\! \tilde{X}: \exists s'\!\in \!\mathcal{L}(\tilde{G})\text{ s.t. }   \tilde{H}_R(s)\!=\!\tilde{H}_R(s')\}.
	\]			
	By Lemma~\ref{lem1} and the construction of $\tilde{G}$, we  have
	\[
	\hat{\tilde{X}}(\tilde{H}_R(s)) \subseteq  \hat{X}(H_R(s)) \times \{N,Y\} . 
	\]  
	Therefore, by the definition of $\tilde{X}_S$, we have that 
	\[
	\hat{\tilde{X}}(\tilde{H}_R(s))\subseteq \tilde{X}_S\Leftrightarrow \hat{X}(H_R(s))\subseteq X_S.
	\]
	
	Hence, $\tilde{G}$ is opaque if and only if $G$ is opaque. \hfill $\qed$
\end{pf}

Based on the above discussion, hereafter, we will only perform analysis based on the augmented system $\tilde{G}$. For the sake of simplicity,  hereafter,  
we still use notations $\hat{X}(\cdot)$to represent the CSE under DIRM $\tilde{X}_r$  for system $\tilde{G}$. 

\subsection{Observations in $\tilde{G}$}
In order to investigate how to estimate states in $\tilde{G}$, we first classify different types of observations. 
Under the DIRM, there are actually two different types of observations: 
\begin{itemize}
	\item 
	the standard instant observation for an observable event: 
	this observation is available when an observable event occurs; \vspace{5pt}
	\item 
	the release of some previous events in $\Sigma_r$: 
	this observation is available when  a state in  $\tilde{X}_{r}$ is reached.
\end{itemize}
Note that the above two types of observations may occur simultaneously when $\tilde{X}_{r}$ is reached by an observable event $\sigma\in \Sigma_o$.

Therefore, events in $\Sigma_r$ is unobservable before reaching a state in $\tilde{X}_{r}$.  
On the other hand, once events in $\Sigma_r$ are released, we know precisely when they occurred. 
To handle this issue, we introduce two observation views: high-level view $\mathcal{O}(\tilde{x})$ and low-level view $\mathcal{O}_L(\tilde{x})$. 

The high-level view assumes that events in $\Sigma_o\cup\Sigma_r$ are always observable. 
From this view, an event $\sigma\in \Sigma$ such that $\tilde{\delta}(\tilde{x},\sigma)!$ is considered as an ``observable" event at augmented state $\tilde{x}\in \tilde{X}$ if 
 (i)	$\sigma\in \Sigma_o\cup\Sigma_r$; or  
 (ii)	$\tilde{\delta}(\tilde{x},\sigma)\in \tilde{X}_r$. 
Therefore, we define 
\[
\mathcal{O}(\tilde{x})
:=\{\sigma\in \Sigma:
\tilde{\delta}(\tilde{x},\sigma)! \wedge [\sigma\in \Sigma_o\cup \Sigma_r\vee \tilde{\delta}(\tilde{x},\sigma)\in \tilde{X}_r]
\}
\] 
as the set of high-level ``observable event'' defined at state $\tilde{x} \in \tilde{X}$  assuming events in $\Sigma_o\cup\Sigma_r$ are observable. 

On the other hand, the low-level view is from the observer/intruder's actual observation considering that we cannot see all events in $\mathcal{O}(\tilde{x})$ immediately since $ \sigma\in \Sigma_r $ is not observable until it is released. Instead, we define 
\[
\mathcal{O}_{L}(\tilde{x})
:=\{\sigma\in \Sigma:
\tilde{\delta}(\tilde{x},\sigma)! \wedge [\sigma\in \Sigma_o\vee \tilde{\delta}(\tilde{x},\sigma)\in \tilde{X}_r]
\}
\]
as the set of low-level ``observable events'' defined at state $\tilde{x} \in \tilde{X}$ under the assumption that only event in $\Sigma_o$ are instantly observable.

Let $ q \in 2^{\tilde{X}} $ be a set of augmented states. In order to capture all observation-equivalent states of $q$, from the high-level view, 
we define $\widetilde{\textsc{UR}}(q)$ as the \emph{unobservable reach w.r.t. $\mathcal{O}(\tilde{x})$}. Unlike the standard unobservation reachable in static observation, this unobservation reach cannot be written in a closed-form since unobservable events are \emph{state-dependent}, e.g., the transition is also observable when $\sigma \in \Sigma_{uo} $ and $\tilde{\delta}(\tilde{x},\sigma)\in \tilde{X}_r$. Alternatively,  we define $\widetilde{\textsc{UR}}(q)$   inductively as follows:
\begin{itemize}
	\item 		
	$q\subseteq \widetilde{\textsc{UR}}(q)$;   \vspace{5pt}
	\item 
	For any $\tilde{x}\in q, \sigma\in \Sigma$ such that $\tilde{\delta}(\tilde{x},\sigma)!$, we have 
	\[
	\tilde{\delta}(\tilde{x},\sigma) \in  \widetilde{\textsc{UR}}(q) 
	\Leftrightarrow
	\sigma \notin \mathcal{O}(\tilde{x}).
	\]
\end{itemize}

Also, from the observer's actual observation, we define 
$\widetilde{\textsc{UR}}_L(q)$  as the \emph{unobservable reach} with respective to the low-level view $\mathcal{O}_L(\tilde{x})$ by assuming one can only observe $\Sigma_o$ instantly and events in $\Sigma_r$ are subjected to the DIRM. Similarly, we define $\widetilde{\textsc{UR}}_L(q)$   inductively as follows:
\begin{itemize}
	\item 		
	$q\subseteq \widetilde{\textsc{UR}}_L(q)$;   \vspace{5pt}
	\item 
	For any $\tilde{x}\in q, \sigma\in \Sigma$ such that $\tilde{\delta}(\tilde{x},\sigma)!$, we have 
	\[
	\tilde{\delta}(\tilde{x},\sigma) \in  \widetilde{\textsc{UR}}_L(q) 
	\Leftrightarrow
	\sigma \notin \mathcal{O}_L(\tilde{x}).
	\]
\end{itemize}

Next, we define how a state estimate $ q \in 2^{\tilde{X}} $ evolves according to different types of observations. 
As we discussed above, the observer may observe an new event occurrence or an information release  or both together.
Therefore,  if an observable event $\sigma\in \Sigma_o\cup \Sigma_r$ from the high-level view (or $\sigma\in \Sigma_o$ from the low-level view) occurs without information release, then the set of states reached immediately is defined by
\[
\widetilde{\textsc{NX}}_\sigma(q)=\{\tilde{x} \in \tilde{X}: \exists \tilde{x}'\in q\text{ s.t. }  \tilde{x}=\tilde{\delta}(\tilde{x}',\sigma)\notin \tilde{X}_r  \}.  
\] 
If an unobservable event triggers an information release, then the set of states reached immediately is defined by 
\[
\widetilde{\textsc{NX}}_r(q)=\{\tilde{x} \in \tilde{X}: \exists \tilde{x}'\!\in\! q,\sigma \!\in\!  \Sigma_{uo}   \text{ s.t. }  \tilde{x}\!=\!\tilde{\delta}(\tilde{x}',\sigma)\!\in\! \tilde{X}_r  \}. 
\]
If an  observable event $\sigma\in \Sigma_o \cup \Sigma_r$ occurs and it also triggers an information release, then the set of states reached immediately is defined by
\[
\widetilde{\textsc{NX}}_{\sigma,r}(q)=\{\tilde{x} \in \tilde{X}: \exists \tilde{x}'\in q   \text{ s.t. }  \tilde{x}=\tilde{\delta}(\tilde{x}',\sigma)\in \tilde{X}_r  \}.
\]

We illustrate the above operators by the following example. 

\begin{example}\upshape
Let us still consider $\tilde{G}$ shown in Figure \ref{ex4-2}. 
For  state set  $ \{0N \}$,   from the high-level view, 
	we have $ \widetilde{\textsc{UR}}(\{0N\})=\{0N, 1N\}$ since only unobservable event $u$ can be executed. 
	However, from the low-level view, 
	event $ h $ is ``unobservable" until reaching a state in $ \tilde{X}_r $. 
	Therefore, we have $ \widetilde{\textsc{UR}}_L(\{0N\})=\{0N, 1N, 2Y, 3Y, 4Y, 5Y\} $. 
	Note that, within this unobservable reach, observable event $a$ can happen from states $3Y,4Y$ and $5Y$; 
	each yields a different observation.  
	For example, from state $5Y$, one can just observe the occurrence of $a$ without information release since the successor state reached, i.e., $7Y$, is not in $\tilde{X}_r$.
	Therefore, the corresponding observable reach  is
	\[
	\widetilde{\textsc{NX}}_{a}(\{0N, 1N, 2Y, 3Y, 4Y, 5Y\})=\{7Y\}  
	\]
	However, from states $3Y$ or $4Y$, the occurrence of $a$  not only generates an instant observation but  also releases some historical information.
	This corresponds to the following observable reach	
	\[
	\widetilde{\textsc{NX}}_{a,r}(\{0N, 1N, 2Y, 3Y, 4Y, 5Y\})=\{6Y,9Y\}. 
	\]
	Note that, from the intruder's point of view, it can still distinguish between state $6Y$ and state $9Y$ since 
	by reaching $6Y$ it will observe $hha$ and by reaching $9Y$ it will observe $ha$.  
	Therefore, only the unobservable and observable reach  still cannot compute the precise state estimate of the system. 
	We need additional information that tracks those unreleased information to correct this state estimate. This will be discussed in the next section. 
		\hfill $\qed$
\end{example}

\section{Verification of Opacity using DIRM-Observer}\label{sec5}

In this section, we discuss how to compute the current-state estimate, which is the key to the verification of current-state opacity, under the dynamic information release mechanism. 
To this end, we propose the DIRM-observer, which is a new DFA
\[
Obs(\tilde{G})= (X_{obs},\Sigma,f,x_{obs,0}), 
\]
where 
\begin{itemize}
	\item 
	$X_{obs} \subseteq \tilde{X}\times 2^{\tilde{X}} \times 2^{\tilde{X}} $ is the set  states; 
	\vspace{5pt}
	\item 
	$x_{obs,0}=  (x_0,  \widetilde{\textsc{UR}}(\{x_0\}),  \widetilde{\textsc{UR}}_L(\{x_0\})  )$ is the initial state; \vspace{5pt}
	\item 
	$f:X_{obs}\times \Sigma \to X_{obs}$ is the partial transition function  defined by: 
	for any $x_{obs}=(\tilde{x},\hat{q},q)\in X_{obs}$ and $\sigma\in \Sigma$, we have \vspace{5pt}
	\begin{itemize}
		\item 
		$f(x_{obs},\sigma)$ is defined if and only if $\tilde{\delta}(\tilde{x},\sigma)$ is defined; \vspace{5pt}
		\item 
		if $f(x_{obs},\sigma)$ is defined, then we have
		$f( (\tilde{x},\hat{q},q),\sigma)=(\tilde{x}',\hat{q}',q')$,  where
		\begin{align}
		\tilde{x}'&\!=\!\tilde{\delta}(\tilde{x},\sigma) \\  
		\hat{q}'&\!= \!
		\left\{\! 
		\begin{array}{l l}
		\hat{q}  \quad
		&\text{if } \sigma \!\notin\! \mathcal{O}(\tilde{x})    \\
		\widetilde{\textsc{UR}}(\hat{q}_{mid} )   \quad
		&\text{if } \sigma \!\in\! \mathcal{O}(\tilde{x})  
		\end{array}
		\right.\label{C2}\\
		q'&\!=\! 
		\left\{\! 
		\begin{array}{l l}
		q\quad
		&\text{if } \sigma \!\notin\! \mathcal{O}_L(\tilde{x})\\
		\widetilde{\textsc{UR}}_L( \widetilde{\textsc{NX}}_\sigma(q) )  \quad
		&\text{if } \sigma \!\in\! \mathcal{O}_L(\tilde{x}) \wedge    \tilde{x}'\notin \tilde{X}_r\\
		\widetilde{\textsc{UR}}_L( \hat{q}_{mid}  )  \quad
		&\text{if } \sigma \!\in\! \mathcal{O}_L(\tilde{x}) \wedge    \tilde{x}'\in \tilde{X}_r  
		\end{array}
		\right. \label{eq:update}
		\end{align}
				where 
		\begin{align}
		\hat{q}_{mid}&\!= \!
		\left\{\! 
		\begin{array}{l l}
		\widetilde{\textsc{NX}}_{\sigma}(\hat{q})    \quad
		&\text{if } \sigma \in \Sigma_{o}\!\cup\! \Sigma_{r} \wedge  \tilde{x}'\notin \tilde{X}_r    \\		
		\widetilde{\textsc{NX}}_{r}(\hat{q})    \quad
		&\text{if } \sigma  \!\in\! \Sigma_{uo}\wedge  \tilde{x}'\in \tilde{X}_r    \\	
		\widetilde{\textsc{NX}}_{\sigma,r}(\hat{q})    \quad
		&\text{if } \sigma  \!\in\! \Sigma_{o}\!\cup\! \Sigma_{r} \wedge  \tilde{x}'\in \tilde{X}_r    \\		
		\end{array}\label{Qmid}
		\right.
		\end{align}
	\end{itemize} 
\end{itemize}   
For the sake of simplicity, we only consider the reachable part of $Obs(\tilde{G})$. 
Also, we note that each state in $Obs(\tilde{G})$ is in the form of $x_{obs}=(\tilde{x},\hat{q},q)$, 
where the first component is an augmented state while the second and the third components are sets of augmented states representing state estimates. 
Then for each state $x_{obs}$, we denote by $X_1(x_{obs}),X_2(x_{obs})$ and $X_3(x_{obs})$ its first, second and third components, respectively. 

The intuition of each component of the DIRM-observer is explained as follows. 
The first component essentially tracks in \emph{actual state} of the augmented system. 
Furthermore, we note that, at each state $x_{obs}=(\tilde{x},\hat{q},q)$, $f(x_{obs},\sigma)$ is defined if and only if $\tilde{\delta}(\tilde{x},\sigma)$ is defined. Therefore, we have $\mathcal{L}(Obs(\tilde{G}))=\mathcal{L}(\tilde{G})=\mathcal{L}(G)$, i.e., $Obs(\tilde{G})$ exactly generates the same language of the original plant. 

The second and third components estimate the state of the system from the high-level view and the low-level view, respectively. 
Specifically, the second component tracks all possible states the system could be in assuming the observer has the knowledge of the occurrences of events in $\Sigma_r$. 
Specifically,  if  $\sigma \!\notin\! \mathcal{O}(\tilde{x})$, i.e., there is no observation upon the occurrence of $\sigma$ from the high-level view, then the state estimate should remain the same. 
If $\sigma \!\in\! \mathcal{O}(\tilde{x})$, 
then we need to first update the estimation upon the observation via observable reach and obtain an intermediate estimate denoted by $\hat{q}_{mid}$. 
Then we compute the unobservable reach of $\hat{q}_{mid}$ to complete the information update. 
As discussed earlier, there are three different types of observation for $\sigma \!\in\! \mathcal{O}(\tilde{x})$. 
Therefore, $\hat{q}_{mid}$ is computed accordingly for each case as follows:
\begin{itemize}
	\item 
	If $\sigma \in \Sigma_{o}\cup \Sigma_{r} $, but $  \tilde{x}'\notin \tilde{X}_r$, 
	i.e., from the high-level view of the second component, one can observe an instant observation in $\Sigma_{o}\cup \Sigma_{r}$ but there is no historical information released,
	then the state estimate is updated by observable reach $\widetilde{\textsc{NX}}_{\sigma}(\hat{q})$ since it can track instant observation $ \sigma $ with no historical information released;\vspace{8pt}
	\item 	
	If $\sigma \in \Sigma_{uo} $, but $  \tilde{x}'\in \tilde{X}_r$, 
	i.e., from the high-level view of  the second component, there is no instant observation in $\Sigma_{o}\cup \Sigma_{r}$ but some historical information released, then the state estimate is updated first by $\widetilde{\textsc{NX}}_{r}(\hat{q}) $ since it tracks no instant observation and there is historical information released; \vspace{8pt}
	\item 	
	If $\sigma \in \Sigma_{o} $, but $  \tilde{x}'\in \tilde{X}_r$, 
	i.e., from the high-level view of the  second component, there are both instant observation in $\Sigma_{o}\cup \Sigma_{r}$ and some historical information released, then the state estimate is updated by observable reach $\widetilde{\textsc{NX}}_{\sigma,r}(\hat{q})$. 
\end{itemize} 
After the above update using observable reach, we take the  unobservable reach $\widetilde{\textsc{UR}}(\hat{q}_{mid} )$ w.r.t. $\mathcal{O}(\tilde{x})$  to capture all states that cannot be distinguished with $\hat{q}_{mid}$. 
Note that, the estimated information in the second component cannot be used directly because the actual observer does not directly observe events in $\Sigma_r$. 
The information in this component is used to construct the third component when information release is triggered. 
This is because, only  in this case, all previous $\sigma\in \Sigma_{r}$ are released and one can then leverage the information in the high-level view to correct the information in the low-level view. 

The third component aims to represent the current-state estimate under the DIRM projection, which is of our main interest. 
The information update of this component is based on the actual low-level view of the intruder. 
Specifically, if $ \sigma \!\notin\! \mathcal{O}_L(\tilde{x})$, i.e., there is no actual observation of the intruder, then the state estimate should remain the same. 
On the other hand, for $ \sigma \!\in\! \mathcal{O}_L(\tilde{x})$, depending on whether or not there is historical information released, we have the following two cases: 
\begin{itemize} 
	\item 
	If  $ \tilde{x}'\!\notin\! \tilde{X}_r$,  
	then the  intruder only has an instant observation $\sigma$ but there is no historical information released. 
	Therefore, the state estimate is updated by observable reach $\widetilde{\textsc{NX}}_{\sigma}(q)$ and the low-level view unobservable reach $\widetilde{\textsc{UR}}_L(q)$ w.r.t. $\mathcal{O}_L(\tilde{x})$; \vspace{8pt}
	\item 
	If  $ \tilde{x}'\!\in\! \tilde{X}_r$, then there is  historical information released. 
	As we explained for the second component, the intermediate estimate $\hat{q}_{mid}$ in the second component actually captures what the intruder knows if it has the high-level view. 
	Since the information has been released, this information becomes available to the low-level intruder. 
	Therefore, instead of taking the observable reach of the previous estimate in the third component, this part is ``switched" to $\hat{q}_{mid}$ to utilize the released information.  Next, since the third component captures the actual state estimate of the intruder, which cannot observe events in $\Sigma_{r}$ before the next information release, 
	we again take the low-level unobservable reach of $\hat{q}_{mid}$ from the intruder's actual point of view to complete the information update. 
\end{itemize}

We use the following example to illustrate the DIRM-observer.
\begin{figure}
	\centering
	\includegraphics[width=0.5\textwidth]{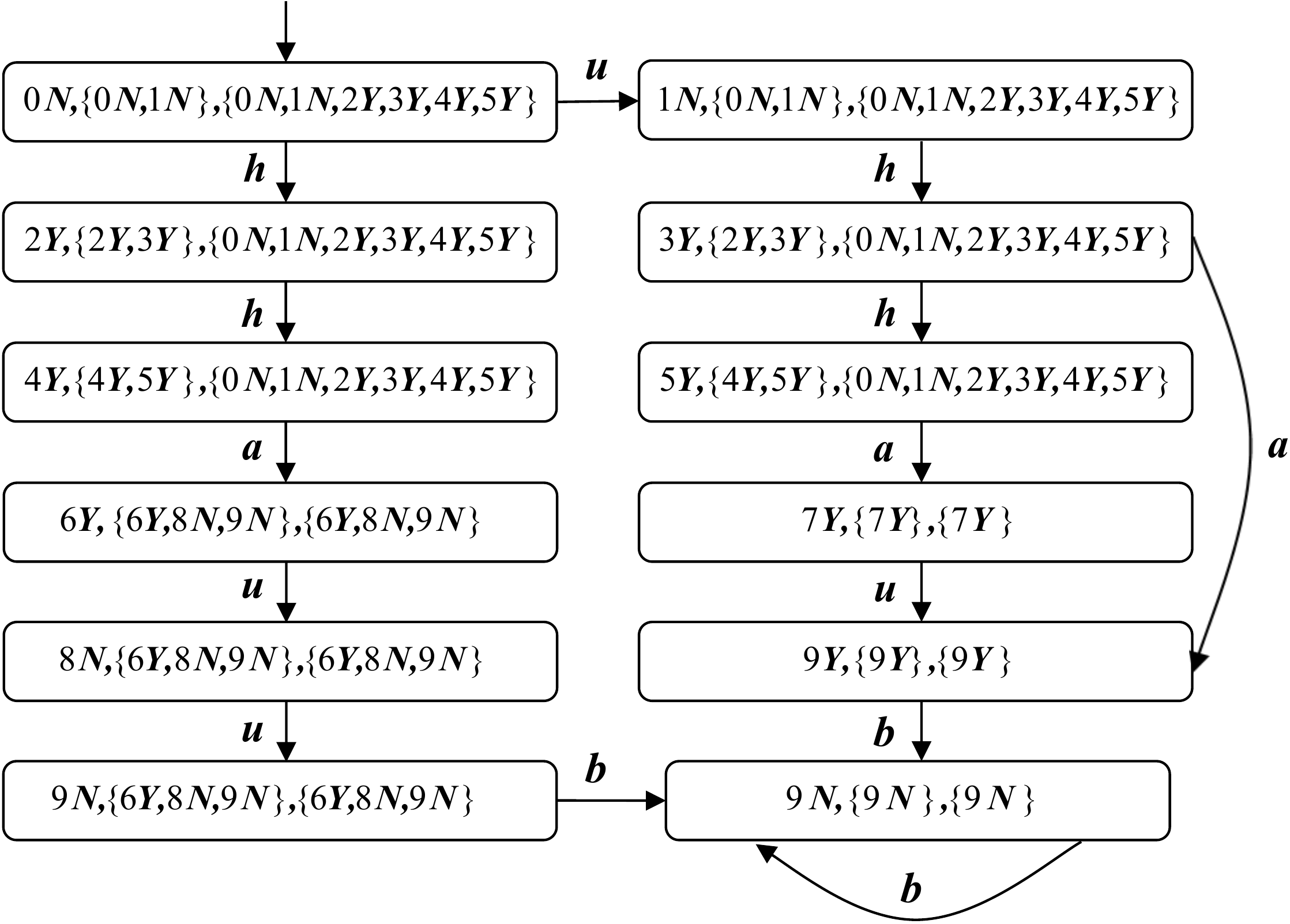} 
	\caption{DIRM-observer $Obs(\tilde{G})$ of $ G $.} 
	\label{ex6obs}  
\end{figure}

\begin{example}\upshape 
Let us still consider DFA $ G =(X,\Sigma,\delta,x_0)$ in Figure~\ref{ex4-1} and its augmented system $\tilde{G}$ in Figure~\ref{ex4-2}. 
The DIRM-observer for $\tilde{G}$ is shown in Figure~\eqref{ex6obs}. 
The initial state is $x_{obs,0}=(0N,\{0N,1N \},$ $\{0N, 1N, 2Y, 3Y, 4Y, 5Y\}     ) $ 
since  $\widetilde{\textsc{UR}}(\{0N\})= \{0N, 1N\} $ and  $\widetilde{\textsc{UR}}_L(\{0N\})= \{0N, 1N, 2Y, 3Y, 4Y, 5Y\} $. 
Therefore, by observing nothing, the state estimate of the intruder is 
$\hat{X}(\{\epsilon\} ) = X_3(x_{obs,0}) = \{0N, 1N, 2Y, 3Y, 4Y, 5Y\}  $. 

For the initial-state, if event $h$ occurs, we have $f(h)=(2Y,\{2Y,3Y \},\{0N, 1N, 2Y, 3Y, 4Y, 5Y\}     )$. 
Note that the first component is updated to $2Y=\tilde{\delta}(0N,h)$ and the second component is updated to $\widetilde{\textsc{UR}}( \widetilde{\textsc{NX}}_{\sigma}(\{0N,1N \}) =\{2Y,3Y \} $ since $h\in \Sigma_r$. 
However, the third component remains unchanged since $h\notin \mathcal{O}_L(0N)=\emptyset$. 
Similarly, when $h$ occurs again, we have $f(hh)=(4Y,\{4Y,5Y \},\{0N, 1N, 2Y,$ $ 3Y, 4Y, 5Y\}     )$ since $h\notin \mathcal{O}_L(2Y)=\emptyset$, i.e., the third component is still not updated.  

Now, let us consider string $hha$, the first component is updated to $6Y=\tilde{\delta}(4Y,a)$ and the second component is updated to $\widetilde{\textsc{UR}}(\widetilde{\textsc{NX}}_{a,r}(  \{4Y,5Y \}  )  )=\{6Y,8N,9N \} $. 
Since $a\in \Sigma_o$ and $6Y\in \tilde{X}_r$, the third component is updated to 
$\widetilde{\textsc{UR}}_L(\hat{q}_{mid}) = \{6Y,8N,9N\}$, 
where $\hat{q}_{mid}= \widetilde{\textsc{NX}}_{a,r}(  \{4Y,5Y \}  )=\{6Y\}$.  
This corresponds to the last case of Equation~\eqref{eq:update}.

For state $(5Y,\{4Y,5Y\},\{0N,1N,2Y,3Y,4Y,5Y\})$, when event $a$ occurs, 
it reaches $(7Y,\{7Y\},$ $\{7Y\})$, where 
since $a\in \Sigma_o$ but $7Y\notin \tilde{X}_r$, the third component is updated according to the second case of Equation~\eqref{eq:update} by 
\begin{align} 
 &\widetilde{\textsc{UR}}_L(\widetilde{\textsc{NX}}_{a}( \{0N,1N,2Y,3Y,4Y,5Y\}   ))\nonumber\\
=&\widetilde{\textsc{UR}}_L(\{7Y\})= \{7Y\}\nonumber 
\end{align}
If event $u$ occurs from $(7Y,\{7Y\},\{7Y\})$, then we move to $(9Y,\{9Y\},\{9Y\})$.  
Since $u\in \Sigma_{uo}$ but $9Y\in \tilde{X}_r$, 
the second component is updated according to the second case of Equation~\eqref{Qmid} by 
$\widetilde{\textsc{UR}}( \widetilde{\textsc{NX}}_{r}(\{7Y\}) ) =\{9Y\}$; 
the third component is updated according to the third case of Equation~\eqref{eq:update} by 
$\widetilde{\textsc{UR}}_L(\hat{q}_{mid})
=\widetilde{\textsc{UR}}_L(\{9Y\}) = \{9Y\}$, 
where $\hat{q}_{mid}= \widetilde{\textsc{NX}}_{r}(\{7Y\})=\{9Y\}$.
\hfill $\qed$
\end{example}

Next, we show the properties  of the DIRM-observer and establish its correctness.
First,    we show that, for a string $s$ and any state in the component $X_2(f(s))$, there exists a string having the same history with  string $s$.  
Its proof is provided in the appendix.
\begin{lemma}\label{lemma:x2h}
Let $ Obs(\tilde{G}) $ be the DIRM-observer of $ \tilde{G} $. Then, for any $ s \in \mathcal{L}(\tilde{G})$, we have 
\[
\forall x\in X_2(f(s)), \exists \ t \in \mathcal{L}(\tilde{G}) : x=\tilde{\delta}(t) \wedge H_R(t)=H_R(s)
\]
\end{lemma}

Using the above result, we can obtain a precise  characterization  for the second component  of the DIRM-observer. 
\begin{proposition}\label{prop:x2}
Let $ Obs(\tilde{G}) $ be the DIRM-observer of $ \tilde{G} $. 
Then for any $s\sigma \in \mathcal{L}(\tilde{G})$ such that $\tilde{\delta}(s\sigma)\!\in\! \tilde{X}_r$,
	we have 
	\begin{align}
	&   \nonumber
	X_2(f(s\sigma)) =&\left\{\! \tilde{\delta}(s'):  
	\begin{array}{c c}
	s'=t\sigma' \omega\!\in\! \mathcal{L}(\tilde{G}), \omega\in \Sigma_{uo}^* \wedge \\
	H_R(t)\!=\!H_R(s) \wedge \tilde{\delta}(t\sigma')\!\in\! \tilde{X}_r\wedge\\
	P_{\Sigma_{o}\cup\Sigma_{r}}(t\sigma')=P_{\Sigma_{o}\cup\Sigma_{r}}(s\sigma)
	\end{array}    \!\! \right\}\nonumber
	\end{align}
\end{proposition}

Also, we characterize the third component of the DIRM-observer by showing that indeed tracks the current-state estimate of system under DIRM projection.
\begin{proposition}\label{prop:main}
	Let $ Obs(\tilde{G}) $ be the DIRM-observer of $ \tilde{G} $. Then, we have 
	\begin{equation}
	\forall s \in \mathcal{L}(\tilde{G}):   X_3(f(s)) = \hat{X}(H_R(s))
	\end{equation}  
\end{proposition}

By putting the above results together,  we finally provide a theorem to show how to verify current-state opacity under DIRM based on the DIRM-observer.
\begin{theorem}\label{the2}
Let $Obs(\tilde{G})$ be the DIRM-observer for system $G$. Then system $G$ is current-state opaque w.r.t.\ $X_S$ and $X_r$ if and only if  
	\begin{align}
	\forall x_{obs}\in X_{obs}: X_3(x_{obs}) \nsubseteq \tilde{X}_S
	\end{align}
\end{theorem}
\begin{pf}
(if) Suppose that  	$\forall x_{obs}\in X_{obs}: X_3(x_{obs}) \nsubseteq \tilde{X}_S$. 
		By Proposition~\ref{prop:main} and $ \mathcal{L}(\tilde{G})=\mathcal{L}(Obs(\tilde{G})) $, we have that $	\forall s \in \mathcal{L}(\tilde{G}):   X_3(f(s)) = \hat{X}(H_R(s))\nsubseteq \tilde{X}_S$.
		Then by Definition~\ref{def:opa-DIRM}, 
		we have that $\tilde{G}$ is  current-state opaque w.r.t.\  $\tilde{X}_S$ and  $\tilde{X}_r$.
		Since $\mathcal{L}(G)= \mathcal{L}(\tilde{G})$, the definition of $\tilde{X}_S$ and  $\tilde{X}_r$, and Theorem~\ref{theo1}, we have that $G$ is  current-state opaque w.r.t.\  $X_S$ and $X_r$.
		
(only if)  By contradiction. Suppose that, for $G$ that is current-state opaque w.r.t.\  secret states $X_S$ and DIRM $X_r$ and the corresponding $\tilde{G}$, $\exists x_{obs}\in X_{obs}: X_3(x_{obs}) \subseteq \tilde{X}_S$.
		Based on Proposition~\ref{prop:main} and the fact that $\mathcal{L}(G)=\mathcal{L}(\tilde{G})=\mathcal{L}(Obs(\tilde{G}))$, we have $\exists s \in \mathcal{L}(\tilde{G}):   X_3(f(s)) = \hat{X}(H_R(s))\subseteq \tilde{X}_S$. By Definition~\ref{def:opa-DIRM} and Theorem~\ref{theo1}, it can conclude that $\exists s \in \mathcal{L}(G): \hat{X}(H_R(s))\subseteq X_S$, which contradicts with the assumption. \hfill $\qed$
\end{pf}

The theorem shows that  current-state opacity can be checked by a reachability search in the DIRM-observer. 
We illustrate Theorem~\ref{the2} by the following example. 

\begin{example}\upshape
Let us still consider DFA $ G =(X,\Sigma,\delta,x_0)$ in Figure~\ref{ex4-1} and its augmented system $\tilde{G}$ in Figure~\ref{ex4-2}. 
The DIRM-observer for $\tilde{G}$ has been shown in Figure~\eqref{ex6obs}, 
where for   $x= (7Y,\{7Y\},$ $\{7Y\})$, we have $X_3(x)=\{ 7Y \}  \subseteq \tilde{X}_S$. 
This state is reached by $uhha$ with $H_R(uhha)=\{\epsilon, a \}$, which is the unique string having this history. 
Therefore, the intruder knows for sure that the system is at secret state $7$ and also, by  Theorem~\ref{the2}, we know that $G$ is not current-state opaque with respect to $X_r$ and $X_S$. 	\hfill $\qed$

\end{example}

\begin{remark}\upshape
Finally, we discuss the complexity of proposed approach for verifying current-state opacity under DIRM. 
First, the augmented system $\tilde{G}$ contains at most $2\cdot|X|$ states which is linear in the size of the original system. 
The DIRM-observer contains at most $|\tilde{X}|\cdot 2^{|\tilde{X}|} \cdot 2^{|\tilde{X}|}$ states and $|\Sigma|\cdot|\tilde{X}|\cdot 2^{|\tilde{X}|} \cdot 2^{|\tilde{X}|}$ transitions. 
Therefore, the overall complexity for verifying current-state opacity is exponential in the number of states in the original system. 
However, this complexity seems to be unavoidable since verifying current-state opacity under the standard natural projection, which is a special case of the DIRM, is already shown to be PSPACE-hard. 
\end{remark}

\section{Case Study on Medical Cloud Computing Services}\label{sec6}

\begin{figure*}
	\centering
	\includegraphics[width=0.75 \textwidth]{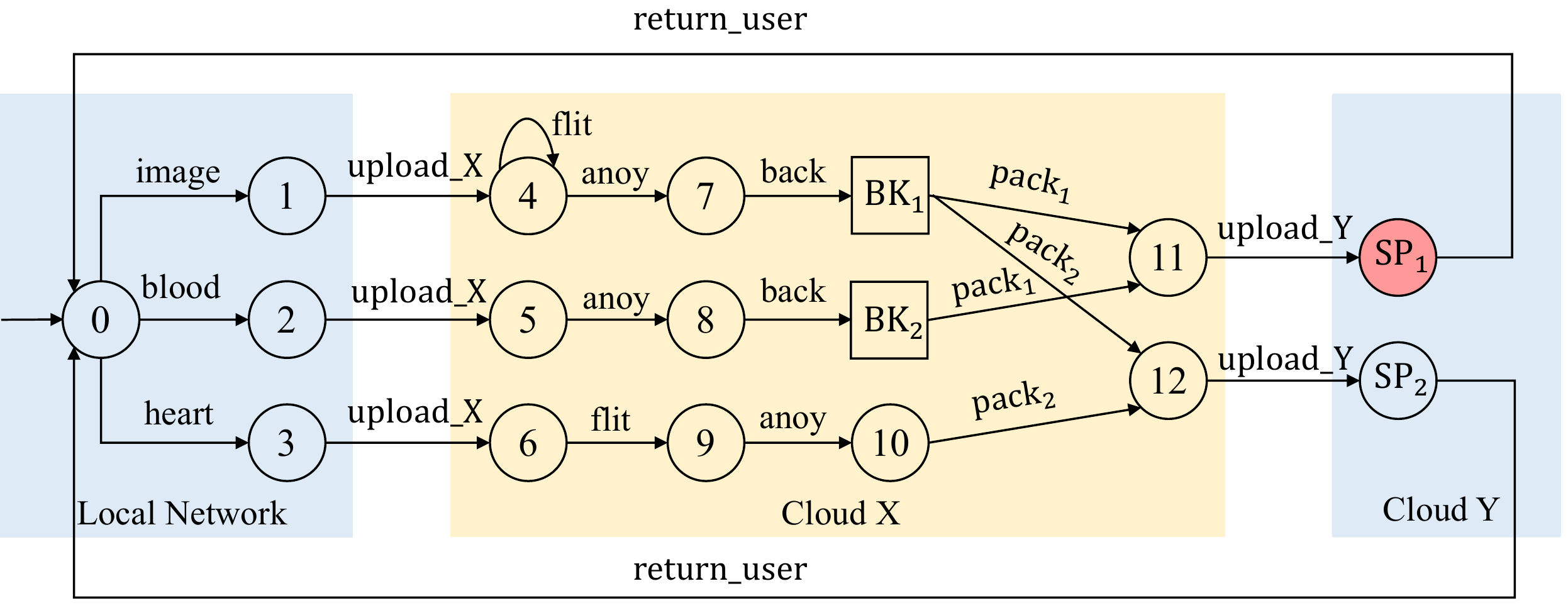} 
	\caption{The operation flow in a cloud based medical service system.} 
	\label{app}  
\end{figure*}

In this section, we apply the proposed framework to a cloud-computing-based medical data processing example. This example is adopted from  \cite{zeng2019quantitative} with some modifications and simplifications.  The reader is referred to \cite{zeng2019quantitative}  for details on how to model medical cloud computing systems using DES models.

Specifically, we consider a scenario, where an user wants to analyze some medical data of patients by cloud computing services.
The entire process involves a local network and two clouds $X$ and $Y$ as shown in Figure \ref{app}; each of them works as follows:
\begin{itemize}
    \item 
    Initially, the user in the local network 
    extracts different health data for later processing. Here, we consider three data extraction operations: 
    \texttt{image} representing the CT image data of the patient, \texttt{blood} representing the blood text data of the patient and \texttt{heart} representing the heart rate sequence data of the patient. 
    Then the extracted data are uploaded to Cloud $X$ for further pre-processing represented by operation $\texttt{upload\_X}$. 
    \medskip
    \item 
    Cloud $X$ aims to pre-process  data received from the local network. 
    Depending on different types of patient data, 
    different pre-processing services are provided, including data filtering operation \texttt{filt} and data anonymization operation  \texttt{anoy}. 
    As shown in   in Figure \ref{app}, 
    image data may need to be filtered for multiple times, 
    while   blood text data does not need to be filtered. 
    The pre-processed data will be packaged to the sender nodes, 
    where the data are sent to service providers $\texttt{SP}_1$ or $\texttt{SP}_2$ in Cloud $Y$ for final process under operation $\texttt{upload\_Y}$. 
    We assume that the image data can be processed by either $\texttt{SP}_1$ or $\texttt{SP}_2$, 
    while the blood data and the heart rate data can only be processed by $\texttt{SP}_1$ and $\texttt{SP}_2$, respectively. 
    Furthermore, before packaging the data, the CT image data and the blood text data also need to be backuped represented by operation 
    \texttt{back}.
    \medskip
	\item 
	Cloud $ Y $ further analyzes the data pre-processed by Cloud $X$ either by service providers $\texttt{SP}_1$ or by $\texttt{SP}_2$ in it. 
	Then the final results will be send 
    back to the user in the local network by operation $\texttt{return\_user}$.
\end{itemize} 
 
Although clouding computing provides a powerful way for processing big local data, one of its main concerns is the privacy issue. Here we consider the following information release scenario. 
We assume that the communication between the local network, Cloud $X$ and Cloud $Y$ are available to the outsider, i.e., we have 
\[
\Sigma_o=\{ \texttt{upload\_X},\texttt{upload\_Y}, \texttt{return\_user} \}
\]
We assume that the operations in within each cloud and the local network are not available directly to the outsider. 
However, Cloud $X$ is subject to the so called \emph{log attacks} \cite{sanamrad2013query,Chu2013security} at the backup locations $\texttt{BK}_i, i=1,2$. 
Specifically, the attacker is able to recover operations that has been executed in the cloud with the right order by checking its logs. 
Therefore, pre-processing operations \texttt{filt} and \texttt{anoy} are events that are initial unobservable but will be released at locations  $\texttt{BK}_i, i=1,2$.   
Formally, we have  
\[
\Sigma_{uo}=\{   \texttt{image}, \texttt{blood}, \texttt{heart}, \texttt{pack}_1, \texttt{pack}_2 \}
\]
and releasable events and the release states are 
\[
\Sigma_{r}=\{   \texttt{filt}, \texttt{anoy} ,\texttt{back} \}\text{ and }X_r=\{\texttt{BK}_1,\texttt{BK}_2\}.
\]
Here we consider a privacy constraint requiring that the intruder should never be able to discover that the user is utilizing service provider $\texttt{SP}_1$ in Cloud $Y$. 
This is because $\texttt{SP}_1$ can provide more detailed analysis than $\texttt{SP}_2$, and therefore, it is used when the pre-processing procedures in Cloud $X$ finds that the patients have high risk of diseases. 
Here,  
we can convert this privacy requirement as a current-state opacity problem by considering 
$X_S=\{\texttt{SP}_1\}$. 
To verify current-state opacity, one can construct the DIRM-observer, which is omited here, and 
we can conclude easily that the entire system is current-state opacity under DIRM. 
For example, for execution 
\[
s= \texttt{blood} \cdot \texttt{upload\_X}  \cdot \texttt{anoy} \cdot\texttt{back} \cdot\texttt{pack}_1\cdot \texttt{upload\_Y} 
\]
that utilizes service provider $\texttt{SP}_1$, there exists another execution 
\[
t= \texttt{image} \cdot \texttt{upload\_X}  \cdot \texttt{anoy} \cdot\texttt{back} \cdot\texttt{pack}_2\cdot \texttt{upload\_Y} 
\]
that utilizes service provider $\texttt{SP}_2$, which does not indicate sensitive information of patients. 

Note that this example cannot be modeled using the standard natural projection. 
This is because whether or not the executions of operations $\texttt{filt}$ and $\texttt{anoy}$ will be discovered by the intruder depends on whether the system will visit release state $\texttt{BK}_i$ in the future.  
Therefore, the proposed DIRM is used to capture the information-flow in this log-attack scenario.

\section{Conclusion}\label{sec7}
In this paper, we proposed a new Orwellian-type observation model called the dynamic information release mechanism that allows to release history information generated dynamically based on the current state.  
A new history-based definition of current-state opacity was proposed that better captures the feature of non-prefix-closed observations, which is the main feature of Orwellian-type observations.   
Then we proposed a new information structure called the DIRM-observer that effectively fuses the released previous information to update the current-state estimate. 
Based on the DIRM-observer, we showed that the current-state opacity verification problem can be effectively solved. 
An illustrative example on medical cloud computing is provided to   illustrate our framework.

There are many interesting future directions under the proposed DIRM framework. 
First, it is interesting to investigate other types of opacity, e.g., initial-state opacity or infinite/$K$-step opacity, under the DIRM. 
Furthermore, in this work, it is assumed that the information release policy is already given (encoded as a state-based function) and we want to verify opacity under a given release policy. 
It is also very interesting to investigate how to synthesize an information release policy that releases historical information as frequently as possible but still preserves opacity.  

\bibliographystyle{plain} 
\bibliography{bibfile}

\begin{thebibliography}{10}

\bibitem{an2020opacity}
L.~An and G.-H. Yang.
\newblock Opacity enforcement for confidential robust control in linear
  cyber-physical systems.
\newblock {\em IEEE Transactions on Automatic Control}, 65(3):1234--1241, 2020.

\bibitem{balun2019opacity}
J.~Balun and T.~Masopust.
\newblock On opacity verification for discrete-event systems.
\newblock {\em arXiv:1912.07314}, 2019.

\bibitem{barcelos2018enforcing}
R.J. Barcelos and J.C. Basilio.
\newblock Enforcing current-state opacity through shuffle in event
  observations.
\newblock {\em IFAC-PapersOnLine}, 51(7):100--105, 2018.

\bibitem{behinaein2019optimal}
B.~Behinaein, F.~Lin, and K.~Rudie.
\newblock Optimal information release for mixed opacity in discrete-event
  systems.
\newblock {\em IEEE Transactions on Automation Science and Engineering},
  16(4):1960--1970, 2019.

\bibitem{hadj2005characterizing}
N.~Ben Hadj-Alouane, S.~Lafrance, F.~Lin, J.~Mullins, and M.~Yeddes.
\newblock Characterizing intransitive noninterference for 3-domain security
  policies with observability.
\newblock {\em IEEE Transactions on AAutomatic Control}, 50(6):920--925, 2005.

\bibitem{hadj2005verification}
N.~Ben Hadj-Alouane, S.~Lafrance, F.~Lin, J.~Mullins, and M.~Yeddes.
\newblock On the verification of intransitive noninterference in mulitlevel
  security.
\newblock {\em IEEE Transactions on Systems, Man, and Cybernetics, Part B
  (Cybernetics)}, 35(5):948--958, 2005.

\bibitem{berard2014verification}
B.~B{\'e}rard and J.~Mullins.
\newblock Verification of information flow properties under rational
  observation.
\newblock {\em arXiv:1409.0871}, 2014.

\bibitem{bryans2008opacity}
J.W. Bryans, M.~Koutny, L.~Mazar{\'e}, and P.~Ryan.
\newblock Opacity generalised to transition systems.
\newblock {\em Internationa Journal of Information Security}, 7(6):421--435,
  2008.

\bibitem{bryans2005modelling}
J.W. Bryans, M.~Koutny, and P.~Ryan.
\newblock Modelling opacity using {P}etri nets.
\newblock {\em Electronic Notes in Theoretical Computer Science}, 121:101--115,
  2005.

\bibitem{Lbook}
C.G. Cassandras and S.~Lafortune.
\newblock {\em Introduction to Discrete Event Systems}.
\newblock Springer, 2nd edition, 2008.

\bibitem{cassez2012synthesis}
F.~Cassez, J.~Dubreil, and H.~Marchand.
\newblock Synthesis of opaque systems with static and dynamic masks.
\newblock {\em Formal Methods in System Design}, 40(1):88--115, 2012.

\bibitem{chen2017quantification}
J.~Chen, M.~Ibrahim, and R.~Kumar.
\newblock Quantification of secrecy in partially observed stochastic discrete
  event systems.
\newblock {\em IEEE Trans.\ Automation Science and Engineering},
  14(1):185--195, 2017.

\bibitem{cong2019line}
X.~Cong, M.P. Fanti, A.M. Mangini, and Z.~Li.
\newblock On-line verification of initial-state opacity by {P}etri nets and
  integer linear programming.
\newblock {\em ISA transactions}, 93:108--114, 2019.

\bibitem{dubreil2010supervisory}
J.~Dubreil, P.~Darondeau, and H.~Marchand.
\newblock Supervisory control for opacity.
\newblock {\em IEEE Trans.\ Automatic Control}, 55(5):1089--1100, 2010.

\bibitem{falcone2015enforcement}
Y.~Falcone and H.~Marchand.
\newblock Enforcement and validation (at runtime) of various notions of
  opacity.
\newblock {\em Discrete Event Dynamic Systems}, 25(4):531--570, 2015.

\bibitem{hadjicostis2020estimation}
C.N. Hadjicostis.
\newblock {\em Estimation and Inference in Discrete Event Systems}.
\newblock Springer, 2020.

\bibitem{jacob2016overview}
R.~Jacob, J.-J. Lesage, and J.-M. Faure.
\newblock Overview of discrete event systems opacity: Models, validation, and
  quantification.
\newblock {\em Annual Rev.\ Control}, 41:135--146, 2016.

\bibitem{ji2018enforcement}
Y.~Ji, Y.-C. Wu, and S.~Lafortune.
\newblock Enforcement of opacity by public and private insertion functions.
\newblock {\em Automatica}, 93:369--378, 2018.

\bibitem{keroglou2016probabilistic}
C.~Keroglou and C.N. Hadjicostis.
\newblock Probabilistic system opacity in discrete event systems.
\newblock {\em Discrete Event Dynamic Systems}, pages 1--26, 2017.

\bibitem{lafortune2018history}
S.~Lafortune, F.~Lin, and C.N. Hadjicostis.
\newblock On the history of diagnosability and opacity in discrete event
  systems.
\newblock {\em Annual Reviews in Control}, 45:257--266, 2018.

\bibitem{lin2011opacity}
F.~Lin.
\newblock Opacity of discrete event systems and its applications.
\newblock {\em Automatica}, 47(3):496--503, 2011.

\bibitem{lin2020information}
F.~Lin, W.~Chen, W.~Wang, and F.~Wang.
\newblock Information control in networked discrete event systems and its
  application to battery management systems.
\newblock {\em Discrete Event Dynamic Systems}, pages 1--26, 2020.

\bibitem{liu2020k}
R.~Liu, L.~Mei, and J.~Lu.
\newblock {$K$}-memory-embedded insertion mechanism for opacity enforcement.
\newblock {\em Systems \& Control Letters}, 145:104785, 2020.

\bibitem{liu2020verification}
S.~Liu and M.~Zamani.
\newblock Verification of approximate opacity via barrier certificates.
\newblock {\em IEEE Control Systems Letters}, 2020.

\bibitem{mohajerani2020compositional}
S.~Mohajerani, Y.~Ji, and S.~Lafortune.
\newblock Compositional and abstraction-based approach for synthesis of edit
  functions for opacity enforcement.
\newblock {\em IEEE Transactions on Automatic Control}, 65(8):3349--3364, 2020.

\bibitem{mohajerani2020ransforming}
S.~Mohajerani and S.~Lafortune.
\newblock Transforming opacity verification to nonblocking verification in
  modular systems.
\newblock {\em IEEE Transactions on Automatic Control}, 65(4):1739--1746, 2020.

\bibitem{mullins2014opacity}
J.~Mullins and M.~Yeddes.
\newblock Opacity with orwellian observers and intransitive non-interference.
\newblock In {\em 12th Int.\ Workshop on Discrete Event Systems}, pages
  344--349, 2014.

\bibitem{ramasubramanian2020notions}
B.~Ramasubramanian, W.R. Cleaveland, and S.~Marcus.
\newblock Notions of centralized and decentralized opacity in linear systems.
\newblock {\em IEEE Transactions on Automatic Control}, 265(4):1442--1455,
  2020.

\bibitem{saadaoui2020current}
I.~Saadaoui, Z.~Li, and N.~Wu.
\newblock Current-state opacity modelling and verification in partially
  observed petri nets.
\newblock {\em Automatica}, 116:108907, 2020.

\bibitem{saboori2011verification}
A.~Saboori and C.N. Hadjicostis.
\newblock Verification of $ k $-step opacity and analysis of its complexity.
\newblock {\em IEEE Trans.\ Automation Science and Engineering}, 8(3):549--559,
  2011.

\bibitem{saboori2012verification}
A.~Saboori and C.N. Hadjicostis.
\newblock Verification of infinite-step opacity and complexity considerations.
\newblock {\em IEEE Trans.\ Automatic Control}, 57(5):1265--1269, 2012.

\bibitem{saboori2013verification}
A.~Saboori and C.N. Hadjicostis.
\newblock Verification of initial-state opacity in security applications of
  discrete event systems.
\newblock {\em Information Sciences}, 246:115--132, 2013.

\bibitem{takai2008formula}
S.~Takai and Y.~Oka.
\newblock A formula for the supremal controllable and opaque sublanguage
  arising in supervisory control.
\newblock {\em SICE J.\ Control, Measu.\ \& Syst.\ Integration}, 1(4):307--311,
  2008.

\bibitem{thorsley2007active}
D.~Thorsley and D.~Teneketzis.
\newblock Active acquisition of information for diagnosis and supervisory
  control of discrete event systems.
\newblock {\em Discrete Event Dynamic Systems}, 17(4):531--583, 2007.

\bibitem{tong2017verification}
Y.~Tong, Z.~Li, C.~Seatzu, and A.~Giua.
\newblock Verification of state-based opacity using {P}etri nets.
\newblock {\em IEEE Trans.\ Automatic Control}, 62(6):2823--2837, 2017.

\bibitem{wang2010optimal}
W.~Wang, S.~Lafortune, A.R. Girard, and F.~Lin.
\newblock Optimal sensor activation for diagnosing discrete event systems.
\newblock {\em Automatica}, 46(7):1165--1175, 2010.

\bibitem{watson2012multi}
P.~Watson.
\newblock A multi-level security model for partitioning workflows over
  federated clouds.
\newblock {\em Journal of Cloud Computing: Advances, Systems and Applications},
  1(1):1--15, 2012.

\bibitem{wu2013comparative}
Y.-C. Wu and S.~Lafortune.
\newblock Comparative analysis of related notions of opacity in centralized and
  coordinated architectures.
\newblock {\em Discrete Event Dynamic Systems}, 23(3):307--339, 2013.

\bibitem{yeddes2016enforcing}
M.~Yeddes.
\newblock Enforcing opacity with orwellian observation.
\newblock In {\em 2016 13th International Workshop on Discrete Event Systems
  (WODES)}, pages 306--312. IEEE, 2016.

\bibitem{yin2017new}
X.~Yin and S.~Lafortune.
\newblock A new approach for the verification of infinite-step and k-step
  opacity using two-way observers.
\newblock {\em Automatica}, 80:162--171, 2017.

\bibitem{yin2019general}
X.~Yin and S.~Lafortune.
\newblock A general approach for optimizing dynamic sensor activation for
  discrete event systems.
\newblock {\em Automatica}, 105:376--383, 2019.

\bibitem{yin2020synthesis}
X.~Yin and S.~Li.
\newblock Synthesis of dynamic masks for infinite-step opacity.
\newblock {\em IEEE Trans.\ Automatic Control}, 65(4):1429--1441, 2020.

\bibitem{yin2019infinite}
X.~Yin, Z.~Li, W.~Wang, and S.~Li.
\newblock Infinite-step opacity and {$K$}-step opacity of stochastic
  discrete-event systems.
\newblock {\em Automatica}, 99:266--274, 2019.

\bibitem{yin2020approximate}
X.~Yin, M.~Zamani, and S.~Liu.
\newblock On approximate opacity of cyber-physical system.
\newblock {\em IEEE Transactions on Automatic Control}, 2020.

\bibitem{zeng2019quantitative}
W.~Zeng and M.~Koutny.
\newblock Quantitative analysis of opacity in cloud computing systems.
\newblock {\em IEEE Transactions on Cloud Computing}, 2019.

\bibitem{zhang2015max}
B.~Zhang, S.~Shu, and F.~Lin.
\newblock Maximum information release while ensuring opacity in discrete event
  systems.
\newblock {\em IEEE Trans.\ Automation Science and Engineering},
  12(4):1067--1079, 2015.

\bibitem{zinck2020enforcing}
G.~Zinck, L.~Ricker, H.~Marchand, and L.~H{\'e}lou{\"e}t.
\newblock Enforcing opacity in modular systems.
\newblock In {\em IFAC World Congress}, 2020.

\end{thebibliography}

\appendix
\textbf{Appendix:  Proofs not contained in main body} 

\textbf{Proof of Lemma~\ref{lemma:x2h}}\\  
	We prove it by induction on the length of $ s $.
	
	{\textit{Induction Basis:}} 
	Suppose that $ |s| = 0 $, i.e., $s = \epsilon$ and $H_R(\epsilon)=\{\epsilon\}$. 
	Then we easily know that $ X_2(f(\epsilon))=\widetilde{\textsc{UR}}(\{x_0\})=H_R(\epsilon)$. Therefore, the induction basis holds.
	
	{\textit{Induction Step:}} 
	Now, let us assume that, for string $ s \in \mathcal{L}(\tilde{G})$ with $|s|=k$, we have $ \forall x\in X_2(f(s)), \exists \ t \in \mathcal{L}(\tilde{G}): x=\tilde{\delta}(t) \wedge H_R(t)=H_R(s) $. Then we consider   $ s\sigma \in \mathcal{L}(\tilde{G})$, $\sigma\in \Sigma$.
	
	For any $x\in X_2(f(s\sigma))$, by the definition of $X_2(f(s))$, we have  $\tilde{\delta}(t)\in X_2(f(s))$, then there exists $t'=t\sigma'\omega \in \mathcal{L}(\tilde{G})$ and $\omega \in \Sigma_{uo}^*$ such that $x=\tilde{\delta}(t')$.
	As the construction of $ X_2 $, we know the events in $\Sigma_{o}\cup\Sigma_{r}$ is observable. Hence, by the definition of the specified observable reach of $ X_2(f(s\sigma)) $ and the induction hypothesis $H_R(t)=H_R(s)$, we have 
	\begin{equation}\label{pro1:main}
	P_{\Sigma_{o}\cup\Sigma_{r}}(t')=P_{\Sigma_{o}\cup\Sigma_{r}}(s\sigma)
	\end{equation}
	\begin{equation}\label{pro1:mainpr}
	P_R(t)=P_R(s)
	\end{equation}
	Next, we consider different cases.
	
	\textit{\textbf{Case 1:}}   $\sigma \in \Sigma_{o} \wedge  \tilde{x}'\in \tilde{X}_r$ \\
	For this case, $ X_2(f(s\sigma))=\widetilde{\textsc{UR}}(\widetilde{\textsc{NX}}_{\sigma,r}(X_2(f(s)) ) $, then we have $\sigma=\sigma'\in \Sigma_{o}$ and $\tilde{\delta}(t\sigma')\in \tilde{X}_r$. Furthermore, the above makes that 
	\begin{equation}\label{pro1:2}
	P_R(s\sigma)=P_{\Sigma_{o}\cup\Sigma_{r}}(s\sigma)\wedge
	P_R(t')=P_R(t\sigma')=P_{\Sigma_{o}\cup\Sigma_{r}}(t\sigma')
	\end{equation}
	By the induction hypothesis, Equations (\ref{pro1:main}) and (\ref{pro1:2}),
	we have 
	$H_R(s)\cup P_R(s\sigma)=H_R(t)\cup P_R(t')$, i.e., $H_R(s\sigma)=H_R(t')$. Therefore, for $\sigma \in \Sigma_{o} \wedge  \tilde{x}'\in \tilde{X}_r$, we have 
	$\forall x\in X_2(f(s\sigma)), \exists \ t' \in \mathcal{L}(\tilde{G}): x=\tilde{\delta}(t') \wedge H_R(t')=H_R(s\sigma)$, i.e., the hypothesis holds in this case.
	
	\textit{\textbf{Case 2:}}   $\sigma \in \Sigma_{o}\cup\Sigma_{r} \wedge  \tilde{x}'\notin \tilde{X}_r$ \\
	For this case, $ X_2(f(s\sigma))=\widetilde{\textsc{UR}}( \widetilde{\textsc{NX}}_{\sigma}(X_2(f(s))) )  $, then we have $\sigma=\sigma'$ and $\tilde{\delta}(t\sigma')\notin \tilde{X}_r$. Furthermore, the above makes that,
	
	when $\sigma \in \Sigma_{r}$,
	\begin{equation}\label{pro1:3}
	P_R(s\sigma)=P_R(s)\wedge
	P_R(t')=P_R(t)
	\end{equation}
	when $\sigma \in \Sigma_{o}$, 
	\begin{equation}\label{pro1:4}
	P_R(s\sigma)=P_R(s)\sigma\wedge
	P_R(t')=P_R(t\sigma')=P_R(t)\sigma'
	\end{equation}
	By the induction hypothesis, Equations (\ref{pro1:mainpr}), (\ref{pro1:3}) or (\ref{pro1:4}),
	we have 
	$H_R(s)\cup P_R(s\sigma)=H_R(t)\cup P_R(t')$, i.e., $H_R(s\sigma)=H_R(t')$. Therefore, the hypothesis holds in this case.

	\textit{\textbf{Case 3:}}   $\sigma \in \Sigma_{uo} \wedge  \tilde{x}'\in \tilde{X}_r$ \\
	For this case, $ X_2(f(s\sigma))=\widetilde{\textsc{UR}}( \widetilde{\textsc{NX}}_{r}(X_2(f(s)))) $, then we have $\sigma, \sigma'\in \Sigma_{uo}$ and $\tilde{\delta}(t\sigma')\in \tilde{X}_r$. Furthermore, the above makes that 
	\begin{equation}\label{pro1:5}
	P_R(s\sigma)=P_{\Sigma_{o}\cup\Sigma_{r}}(s\sigma)\wedge
	P_R(t')=P_R(t\sigma')=P_{\Sigma_{o}\cup\Sigma_{r}}(t\sigma')
	\end{equation}
	By the induction hypothesis, Equations (\ref{pro1:main}) and (\ref{pro1:5}),
	we have 
	$H_R(s)\cup P_R(s\sigma)=H_R(t)\cup P_R(t')$, i.e., $H_R(s\sigma)=H_R(t')$. Therefore, the hypothesis holds in this case.
	
	\textit{\textbf{Case 4:}}   $\sigma \in \Sigma_{uo} \wedge  \tilde{x}'\notin \tilde{X}_r$ \\
	For this case, we have $ X_2(f(s\sigma))=X_2(f(s))) $,  $t'=t\omega\in \mathcal{L}(\tilde{G})$ and $\omega \in \Sigma_{uo}^*$. Further, we easily have that,
	$H_R(s\sigma)=H_R(s)=H_R(t)=H_R(t')$. Therefore, the hypothesis holds in this case.
	
	Based the above cases, this proposition holds.
	\hfill $\qed$

\textbf{Proof of Proposition~\ref{prop:x2}}\\ 
	($\subseteq$) 
	According to the lemma \ref{lemma:x2h} and $\tilde{\delta}(s\sigma)\!\in\! \tilde{X}_r$, we have that $\forall x\in X_2(f(s\sigma)), \exists \ t\sigma'\in \mathcal{L}(\tilde{G}) : x=\tilde{\delta}(t\sigma') \wedge H_R(t\sigma')=H_R(s\sigma)$ and $P_{\Sigma_{o}\cup\Sigma_{r}}(t\sigma')=P_{\Sigma_{o}\cup\Sigma_{r}}(s\sigma)$. Also, for $t\sigma'$, we have $ t'\omega=t\sigma' $, $\omega\in \Sigma_{uo}^*$ such that $ \tilde{\delta}(t')\in \tilde{X}_r $. Therefore, the right set includes the left set.
	
	($\supseteq$) By contradiction. Suppose that there exists string $ t\sigma' \in \mathcal{L}(\tilde{G})$ which satisfies the right conditions, but $\tilde{\delta}(t\sigma')\notin X_2(f(s\sigma))$. Since the specified observable reach of the definition of components $ X_2 $ in $ Obs(\tilde{G}) $ and the right conditions, for $t$, there only can be $\tilde{\delta}(t)\notin X_2(f(s))$ and $P_{\Sigma_{o}\cup\Sigma_{r}}(t)=P_{\Sigma_{o}\cup\Sigma_{r}}(s)$. Similarly, we inductively deduce to that there must be $ t'\in \Sigma_{uo}^*\cap \overline{  \{t\}  } $ such that $\tilde{\delta}(t')\notin X_2(f(\epsilon))$. It contradicts with the initial states definition of $X_2$, since $ \tilde{\delta}(t')$ must be in $\widetilde{\textsc{UR}}(\{x_0\})$, but $X_2(f(\epsilon))=\widetilde{\textsc{UR}}(\{x_0\})$. Therefore, any state in the right also includes in the left.	\hfill $\qed$

 \textbf{Proof of Proposition~\ref{prop:main}}\\ 
	We prove it by induction on the length of $ s $.
	
	{\textit{Induction Basis:}} 
	Suppose that $ |s| = 0 $, i.e., $s = \epsilon$ and $H_R(\epsilon)=\{\epsilon\}$. 
	Then we know that
	\begin{align}
	&  \hat{X}(\{\epsilon\}) \label{initial}\\
	=& \left\{\!
	\tilde{\delta}(s')\in \tilde{X}: 
	s'\in \mathcal{L}(\tilde{G}) \wedge  H_R(s')=\{\epsilon\} 
	\right\}\nonumber\\
	=&\left\{\!
	\tilde{\delta}(s')\in \tilde{X}:  s'\in \mathcal{L}(\tilde{G}) \wedge 
	(\forall s''\in \overline{\{s'\} } )[P_R(s'')=\epsilon] 
	\right\}\nonumber\\
	=&\left\{\!
	\tilde{\delta}(s')\in \tilde{X}:\!\!\!\!\!\!\!\!
	\begin{array}{ c c}
	& s'=\sigma_1\cdots\sigma_n \in \mathcal{L}(\tilde{G}) \wedge \\
	&(\forall i\leq n)  [\sigma_{i} \notin \mathcal{O}_L(   \tilde{\delta}(\sigma_1\cdots\sigma_{i-1})  )]
	\end{array}
	\right\}\nonumber
	\end{align}
	
	Since $\tilde{x}_0$ is included in both $\hat{X}(\{\epsilon\})$ and $\widetilde{\textsc{UR}}_L(\{\tilde{x}_0\})$, 
	by a simple inductive argument according to Equation~\eqref{initial} and the definition of $\widetilde{\textsc{UR}}_L$, one can easily conclude that 
	$\hat{X}(\{\epsilon\})= \widetilde{\textsc{UR}}_L(\{\tilde{x}_0\})$, i.e., the induction basis holds.
	
   {\textit{Induction Step:}} 
	Now, let us assume that, for string $ s \in \mathcal{L}(\tilde{G})$ with $|s|=k$, we have $ X_3(f(s))= \hat{X}(H_R(s)) $.
	Then we consider   $ s\sigma \in \mathcal{L}(\tilde{G})$, where $\sigma\in \Sigma$, for the following two cases.   
	
	\textit{\textbf{Case 1:}}   $\tilde{\delta}(s\sigma)\notin \tilde{X}_r $. \\
	
	For this case, we have 
	$H_R(s\sigma)=H_R(s)\cup \{P_R(s\sigma) \}$. 
	If $ \sigma \in \Sigma_{o}$, then
	\begin{align*}
	& \hat{X}(H_R(s\sigma))  \\
	= & \left\{\! \tilde{\delta}(s') :  s'\!\in\! \mathcal{L}(\tilde{G}) \wedge  H_R(s')\!=\!H_R(s) \!\cup\! \{P_R(s)\sigma \} \right\}\nonumber\\
	= & \left\{\! \tilde{\delta}(s') :  
	\begin{array}{c c}
	s'=t\sigma \sigma_1 \dots \sigma_n \!\in\! \mathcal{L}(\tilde{G})  \wedge \\
	H_R(t)\!=\!H_R(s) \wedge \tilde{\delta}(t\sigma)\!\notin\! \tilde{X}_r\wedge\\
	(\forall i\leq n )  [\sigma_{i} \notin \mathcal{O}_L(   \tilde{\delta}(t\sigma \sigma_1\cdots\sigma_{i-1})  )]
	\end{array}     \right\}\nonumber
	\\
	= & \left\{\! \tilde{\delta}(x,\sigma w) :  \!\!
	\begin{array}{c c}
	w\!=\! \sigma_1 \cdots \sigma_n\wedge
	\\
	x\!\in\! \hat{X}(H_R(s) ) \wedge  \tilde{\delta}(x,\sigma)\!\notin\! \tilde{X}_r\wedge\\
	\forall i\!\leq\! n :  
	\sigma_{i} \notin \mathcal{O}_L(\tilde{\delta}(x,\sigma \sigma_1\cdots\sigma_{i-1})  ) 
	\end{array}    \!\! \right\}\nonumber
	\\
	= & \left\{\! \tilde{\delta}(x',  w) :  \!\!
	\begin{array}{c c}
	w\!=\! \sigma_1 \cdots \sigma_n\wedge\\
	x'\in \widetilde{\textsc{NX}}_{\sigma}( \hat{X}(H_R(s) ) ) \wedge \\
	\forall i\!\leq\! n :  \sigma_{i} \notin \mathcal{O}_L(\tilde{\delta}(x',  \sigma_1\cdots\sigma_{i-1})  )
	\end{array}    \!\! \right\}\nonumber\\
	= & \widetilde{\textsc{UR}}_L(\widetilde{\textsc{NX}}_{\sigma}( \hat{X}(H_R(s) ) ))
	\end{align*}
	Still, by the induction hypothesis and the update function for the third component in Equation~\eqref{eq:update}, we have 
	\begin{align}
	X_3(f(s\sigma))&= \widetilde{\textsc{UR}}_L(\widetilde{\textsc{NX}}_{\sigma}( X_3(f(s))   )) \\
	&= \widetilde{\textsc{UR}}_L(\widetilde{\textsc{NX}}_{\sigma}( \hat{X}(H_R(s) ) )) 
	=    \hat{X}(H_R(s\sigma) ) \nonumber
	\end{align}
	If $ \sigma \notin \Sigma_{o} $, then we have 
	$H_R(s\sigma)=H_R(s)$, 
	which means that $\hat{X}(H_R(s\sigma))=\hat{X}(H_R(s)) $. 
	Therefore, we also have
	\begin{equation}
	X_3(f(s\sigma))\!=\!  X_3(f(s)) \!=\! \hat{X}(H_R(s)) \!=\! \hat{X}(H_R(s\sigma) )
	\end{equation}
	
	\textit{\textbf{Case 2:}} 	  $  \tilde{\delta}(s\sigma)\in \tilde{X}_r $.\\
	For this case, we have 
	$H_R(s\sigma)=H_R(s)\cup \{P_{\Sigma_{o}\cup\Sigma_{r}}(s\sigma) \}$. 
	Therefore, we have
	\begin{align}
	& \hat{X}(H_R(s\sigma))  \nonumber
	\\
	= & \left\{\! 
	\tilde{\delta}(s'):  s'\!\in\! \mathcal{L}(\tilde{G}) \wedge  H_R(s')\!=\!H_R(s)\!\cup\! \{P_{\Sigma_{o} \cup \Sigma_{r}}(s\sigma) \}
	\right\}\nonumber
	\\
	= & \left\{\! \tilde{\delta}(s'):  
	\begin{array}{c c}
	s'=t\sigma' \sigma_1 \dots \sigma_n \!\in\! \mathcal{L}(\tilde{G})  \wedge \\
	H_R(t)\!=\!H_R(s) \wedge \tilde{\delta}(t\sigma')\!\in\! \tilde{X}_r\wedge\\
	P_{\Sigma_{o}\cup\Sigma_{r}}(t\sigma')=P_{\Sigma_{o}\cup\Sigma_{r}}(s\sigma)\wedge\\
	\forall i\leq n : \sigma_{i} \notin \mathcal{O}_L(   \tilde{\delta}(t\sigma' \sigma_1\cdots\sigma_{i-1})  ) 
	\end{array}     \right\}\nonumber
	\end{align}
	Since $\widetilde{\textsc{UR}}(q) \subseteq \widetilde{\textsc{UR}}_L(q)$, then
	by proposition \ref{prop:x2}, we further have
\begin{align*}
\hat{X}(H_R(s\sigma)) = \widetilde{\textsc{UR}}_L(X_2(f(s\sigma)))=\widetilde{\textsc{UR}}_L(\hat{q}_{mid})
\end{align*}
	Then, by the induction hypothesis and the update function for the third component in Equation~\eqref{eq:update}, we have 
	\begin{align*}
	\hat{X}(H_R(s\sigma)) = X_3(f(s\sigma))
	\end{align*}
This completes the proof since $  X_3(f(s)) = \hat{X}(H_R(s))$ holds for each case.
\hfill $\qed$ 
\end{document}